\documentclass[acmsmall]{acmart}

\AtBeginDocument{%
  \providecommand\BibTeX{{%
    \normalfont B\kern-0.5em{\scshape i\kern-0.25em b}\kern-0.8em\TeX}}}

\usepackage[utf8]{inputenc}
\usepackage[T1]{fontenc}  
\usepackage[normalem]{ulem}
\usepackage{amsmath}
\usepackage{bbm}
\usepackage{xcolor}
\usepackage{bbding}

\theoremstyle{definition}
\usepackage{natbib}
\usepackage{graphicx}
\usepackage{subfigure}
\usepackage[skip=-1pt]{caption}
\usepackage{amsmath}
\usepackage{makecell}
\usepackage{amsfonts}
\usepackage{hyperref}
\usepackage{enumitem}
\usepackage{booktabs}
\usepackage{multirow}
\usepackage{mathtools}
\usepackage{multicol}
\usepackage{amsmath, bm}
\usepackage{stmaryrd}
\usepackage{longtable}
\usepackage{float}
\DeclareMathOperator*{\argmin}{argmin}
\DeclareMathOperator*{\argmax}{argmax}
\usepackage[linesnumbered, ruled,vlined]{algorithm2e}
\let\oldnl\nl
\newcommand{\nonl}{\renewcommand{\nl}{\let\nl\oldnl}}
\usepackage{longtable}

\newcommand{\eg}{\emph{e.g.}}

\newcommand{\new}[1]{\textcolor{black}{#1}}

\renewcommand\footnotetextcopyrightpermission[1]{} 
\makeatletter
\def\subsubsection{\@startsection{subsubsection}{3}%
  \z@{.3\linespacing\@plus.7\linespacing}{.1\linespacing}%
  {\normalfont\bfseries}}
\makeatother

\graphicspath{ {./fig/}  }

\begin{document}

\title{\new{A Multi-objective Optimization Framework for Multi-stakeholder Fairness-aware Recommendation}}

\author{Haolun Wu}
\authornote{Corresponding Author.}
\email{haolun.wu@mail.mcgill.ca}
\orcid{0000-0002-8206-4969}
\affiliation{%
  \institution{School of Computer Science, McGill University}
  \streetaddress{845 Sherbrooke St W}
  \city{Montreal}
  \state{Quebec}
  \country{Canada}
  \postcode{H3A 0G4}
}

\author{Chen Ma}
\authornotemark[1]
\email{chenma@cityu.edu.hk}
\affiliation{%
  \institution{Department of Computer Science, City University of Hong Kong}
  \streetaddress{83 Tat Chee Avenue, Y6302, Yeung Kin Man Academic Building, Kowloon Tong}
  \country{Hong Kong SAR}
  }

\author{Bhaskar Mitra}
\email{bhaskar.mitra@microsoft.com}
\affiliation{%
  \institution{Microsoft Research}
  \streetaddress{2000 McGill College Ave}
  \city{Montreal}
  \state{Quebec}
  \country{Canada}
  \postcode{H3A 3H3}}

\author{Fernando Diaz}
\email{diazf@acm.org}
\affiliation{%
  \institution{Google, Canadian CIFAR AI Chair}
  \streetaddress{1253 McGill College Ave}
  \city{Montreal}
  \state{Quebec}
  \country{Canada}
  \postcode{H3B 2Y5}}

\author{Xue Liu}
\email{xueliu@cs.mcgill.ca}
\affiliation{%
  \institution{School of Computer Science, McGill University}
  \streetaddress{845 Sherbrooke St W}
  \city{Montreal}
  \state{Quebec}
  \country{Canada}
  \postcode{H3A 0G4}
}

\renewcommand{\shortauthors}{Wu and Ma, et al.}

\begin{abstract}
Nowadays, most online services are hosted on multi-stakeholder marketplaces, where consumers and producers may have different objectives. Conventional recommendation systems, however, mainly focus on maximizing consumers' satisfaction by recommending the most relevant items to each individual. This may result in unfair exposure of items, thus jeopardizing producer benefits. Additionally, they do not care whether consumers from diverse demographic groups are equally satisfied. To address these limitations, we propose a multi-objective optimization framework for fairness-aware recommendation, \textit{Multi-FR}, that adaptively balances accuracy and fairness for various stakeholders with Pareto optimality guarantee.
We first propose four fairness constraints on consumers and producers.
In order to train the whole framework in an end-to-end way, we utilize the smooth rank and stochastic ranking policy to make these fairness criteria differentiable and friendly to back-propagation.
Then, we adopt the multiple gradient descent algorithm to generate a Pareto set of solutions, from which the most appropriate one is selected by the Least Misery Strategy. 
The experimental results demonstrate that \textit{Multi-FR} largely improves recommendation fairness on multiple stakeholders over the state-of-the-art approaches while maintaining almost the same recommendation accuracy.
The training efficiency study confirms our model's ability to simultaneously optimize different fairness constraints for many stakeholders efficiently.
\end{abstract}


\maketitle
\thispagestyle{fancy}

\section{Introduction and Motivation}
%
%
When viewed from a sociotechnical lens, conventional deployed machine learning systems demonstrate a range of socially problematic behaviors, including algorithmic bias and misinformation~\cite{xiangnan_bias_survey}. Multisided recommendation systems in marketplaces, content-distribution networks, and match-making platforms require reasoning about the potential impact on several populations of stakeholders with potentially disparate and possibly conflicting objectives. As such, the potential societal implications of a recommendation algorithm must balance multiple objectives across these groups.   

Two of the most important stakeholders in these multisided systems are (\romannumeral1) producers who provide goods and services, and (\romannumeral2) consumers who purchase them. When a recommendation system systematically overlooks the utility of certain historically disadvantaged groups, inequity can be exacerbated for producers and consumers. We take the movie recommendation platform as an example, where young children are likely to be a minority of consumers and they favor cartoons. If the system is solely concerned with the utility of adults in order to maximize revenues, this may have a detrimental effect on the satisfaction of young children, and hence on the utility of cartoon producers. In this circumstance, the cartoons receive insufficient exposure, and their creators may leave the platform as a result of poor earnings. Such an unbalanced market will work against those adults' utility if they ever wish to watch cartoons someday, thus degrading the overall experience on the platform. 

%
%
Unfortunately, existing approaches for recommendation systems/platforms are stuck in the aforementioned issue. First, conventional accuracy-centric approaches employ various data-driven models~\cite{Koren:2009:MFT:1608565.1608614, 10.5555/3015812.3015891, Desrosiers11acomprehensive, DBLP:journals/corr/abs-2101-04852, DBLP:conf/sigir/SunZGGTHMC20, DBLP:conf/kdd/MaMZTLC20} to estimate the relevance scores of the consumer-product pair, and then recommend the top-$K$ most relevant products to the corresponding consumers. 
However, these approaches can create a huge disparity in the exposure of the producers due to the ``superstar economics'' \cite{MehrotraMBL018CounterTradeoff, baranchuk2011economics}, which is unfair for the producers and also harms the health of the marketplace. Second, many approaches merely address either the consumer-sided fairness~\cite{GeLGXLZP0GOZ21_longtermfairness, racial, LambrechtT19_gender, DBLP:conf/www/SerbosQMPT17} or the producer-sided fairness~\cite{asia:equity-of-attention,pfair_pop, pfair_incremental, DBLP:conf/kdd/SinghJ18, singh:fair-pg-rank, ZehlikeB0HMB17_FAIR}, 
but neglect the fairness on the other side. This is not desirable, as producers and consumers are both indispensable in these marketplaces. 
Third, several approaches~\cite{PatroBGGC20FairRec, Wu0XT21TFROM} consider the fairness on both sides merely at an individual level (e.g., envy-freeness), but ignore the sensitive attributes (e.g., age, gender, popularity) at a group level. 
We regard this as sub-optimal, since the society strives to not overlook the utility of minority groups in the real world.
Thus, the sensitive attributes should be taken into consideration.

Apart from the limitation that no prior works have attempted to model both the consumer-sided and producer-sided fairness at a group level concurrently in a recommendation framework, another disadvantage is that prior approaches~\cite{MehrotraMBL018CounterTradeoff, suhr2019TwosidedShare} on fairness-aware recommendation generally require fine-tuning weights for multiple objectives in order to optimize the overall objective, which is tedious and cannot guarantee a satisfactory final solution. When involving multiple objectives, one optimal solution is that no objective can be further improved without impairing the others. This optimality is widely recognized and is referred to as the \textit{Pareto optimality}~\cite{Pareto_def}. Existing approaches for optimizing multiple objectives can be broadly classified into two categories: (\romannumeral1) heuristic search and (\romannumeral2) scalarization (weighted summation). While multi-objective evolutionary algorithms are frequently used in heuristic search, they ensure that the emerging solutions are not dominated by each other (but can still be dominated by Pareto optimal solutions)~\cite{DBLP:conf/ppsn/KimHMW04}. Thus they cannot guarantee Pareto optimality. The scalarization method converts multiple objectives into a single objective with weighted summation and can achieve Pareto optimal solutions with proper scalarization~\cite{GHANEKANAFI20157483}. However, existing approaches that manually adjust the scaling factors for each objective frequently fail to satisfy the necessary conditions for Pareto optimality. 
As a result, it is still a challenge for current methods in achieving Pareto optimality when optimizing multiple fairness objectives.

%
%

\begin{table}[t]
\centering
\caption{\new{Properties of different recommendation approaches. Our proposed method, Multi-FR, can model both the consumer-sided fairness and producer-sided fairness jointly in one recommendation framework while incorporating sensitive attributes. Our fairness metrics are also differentiable and friendly for back-propagation. Most significantly, our strategy is theoretically guaranteed to be Pareto optimal with respect to all objectives and does not need handcraft tuning on the scaling factors.}}
\vspace{0.2cm}
\label{tab:approach-comparison}
\resizebox{1.0\textwidth}{!}{
\begin{tabular}{c|cccccc}
\toprule
Approaches \& Collections & \makecell{Consumer-sided\\Fairness} & \makecell{Producer-sided\\Fairness} & \makecell{Using Sensitive \\Attributes} &
\makecell{Adaptive Factor\\ Learning} & \makecell{Differentiable\\Metrics} & \makecell{Pareto \\ Optimal}\\
\midrule
\makecell{Accuracy-centric Policy\\ ~\cite{Koren:2009:MFT:1608565.1608614, 10.5555/3015812.3015891, Desrosiers11acomprehensive, DBLP:journals/corr/abs-2101-04852, DBLP:conf/sigir/SunZGGTHMC20, DBLP:conf/kdd/MaMZTLC20}} & \XSolidBrush & \XSolidBrush & \textbf{N.A.} & \textbf{N.A.}& \XSolidBrush & \textbf{N.A.}\\
\makecell{Consumer-sided Fairness\\~\cite{GeLGXLZP0GOZ21_longtermfairness, racial, LambrechtT19_gender, DBLP:conf/www/SerbosQMPT17}} & \CheckmarkBold & \XSolidBrush & \CheckmarkBold/\XSolidBrush & \XSolidBrush& \XSolidBrush & \XSolidBrush\\
\makecell{Producer-sided Fairness\\~\cite{asia:equity-of-attention,pfair_pop, pfair_incremental, DBLP:conf/kdd/SinghJ18, singh:fair-pg-rank, ZehlikeB0HMB17_FAIR, MehrotraMBL018CounterTradeoff}} & \XSolidBrush & \CheckmarkBold & \CheckmarkBold/\XSolidBrush & \XSolidBrush& \XSolidBrush & \XSolidBrush\\
\makecell{Two-sided Fairness  \\~\cite{PatroBGGC20FairRec, suhr2019TwosidedShare, Wu0XT21TFROM}} & \CheckmarkBold & \CheckmarkBold & \CheckmarkBold/\XSolidBrush & \XSolidBrush& \XSolidBrush & \XSolidBrush\\
\midrule
\textbf{Multi-FR (Ours)} & \CheckmarkBold & \CheckmarkBold & \CheckmarkBold & \CheckmarkBold& \CheckmarkBold & \CheckmarkBold\\
\bottomrule
\end{tabular}}
\end{table}

To address the aforementioned problems, we treat the multi-stakeholder fairness-aware recommendation as a multi-objective optimization (MOO) task and propose a scalable framework, namely \textit{Multi-FR}. 
We summarize the differences between prior works and our work in Table~\ref{tab:approach-comparison}. 
\new{As shown, our work is the first one that satisfies all the properties.}
Specifically, \new{we propose a method to differentiate} four fairness metrics on the consumer side and the producer side, taking into account different attributes within this framework. 
\new{Thus, the fairness constraints can be directly optimized through back-propagation.}
\new{We adopt the weighted summation to combine all these fairness constraints as well as the objective for the recommendation accuracy into one unified framework. }
Thereafter, the multiple gradient descent algorithm (MGDA) along with the Frank-Wolfe Solver~\cite{Frank-Wolf, SenerK18MTLasMOO} are utilized to generate scaling factors regarding these fairness and recommendation objectives for satisfying the necessary conditions of Pareto optimality.
It is worthy to notice that the scaling factors can be adaptively updated during the training procedure without handcraft tuning and the whole framework is trained in an end-to-end way. 
Finally, the most appropriate solution is selected by the Least Misery Strategy~\cite{DBLP:journals/mta/PessemierDM14, LinCPSXSZOJ19ParetoRecsys}. 
The experimental results on three public datasets indicate that our method can well balance the recommendation quality and fairness, and significantly outperform other state-of-the-art fairness-aware recommendation methods on all the fairness metrics.

To summarize, our contributions are:
\begin{itemize}[leftmargin=*]
    \item We propose a general fairness-aware recommendation framework with the multi-objective optimization, \textit{Multi-FR}, which jointly optimizes accuracy and fairness for consumers and producers in an end-to-end way. \new{The final solution is guaranteed to be Pareto optimal theoretically.}
    \item We leverage the multiple gradient descent algorithm with the Frank-Wolfe Solver to guarantee that the scaling factors satisfy the necessary conditions of Pareto optimality. Furthermore, these scaling factors are updated adaptively throughout the training procedure, eliminating the need of manual-crafted search.
    \item \new{We propose a method for differentiating the fairness criteria on both the consumer and producer sides through utilizing the smooth rank and stochastic ranking policy, so that these fairness constraints can be optimized directly and are friendly to back-propagation.}
    \item Extensive experimental results on three public datasets comparing with three state-of-the-art fairness-aware recommendation approaches demonstrate that \textit{Multi-FR} largely improves the recommendation fairness with little drop in terms of the recommendation quality. Further analysis indicates the capability of \textit{Multi-FR} in terms of optimizing multiple fairness objectives concurrently with efficiency.
\end{itemize}

\section{Related Work}
In this section, we provide a summary regarding the related studies from the following three aspects: (\romannumeral1) the definition of fairness in recommendation systems, (\romannumeral2) approaches to achieving fairness, and (\romannumeral3) recommendation with multiple objectives.
\new{We close this section by highlighting the novelty and difference of our work compared to prior methods.}

\subsection{Definition of Fairness in Recommendation}
Prior works on fairness in recommendation, from the perspective of stakeholders, consider algorithmic effects on consumers (i.e. users who seek content) and producers (i.e. users who provide content), independently or together.

On the consumer side, fairness refers to systematic differential performance \cite{mehrotra:demographics-of-search} across consumers. 
Some works focus on the individual fairness that ensures similar individuals are treated similarly~\cite{DworkHPRZ12FairnessThroughAwareness}. Other works define fairness on a group level and aim to make the system provide
comparable quality of service or utility to consumers within different demographic groups (e.g., gender, race, age)~\cite{ekstrand2018all}. \citet{chaney:recsys-feedback-loops} demonstrate, through simulation, that feedback loops inherent in the production system can exacerbate unfairness and homogenize recommendation. \citet{yao:beyond-parity} demonstrate that these issues can be addressed by introducing fairness constraints during the training process.

Other works focus on fairness on the producer side, whose fairness can be defined as the systematic differential exposure \cite{asia:equity-of-attention,singh:fair-pg-rank,DBLP:conf/cikm/DiazMEBC20, MehrotraMBL018CounterTradeoff} across content producers and, most often, groups of producers (e.g. grouped by genre or popularity). For instance, \citet{ekstrand:book-provider-fairness} find that standard recommendation algorithms may result in certain demographic groups of producers being over- or under-represented in recommendation decisions. \citet{beutel:pairwise-fairness} demonstrate that these issues can be addressed in production systems by defining pairwise fairness objectives and introducing them as learning objectives.

Joint satisfaction for consumer-sided fairness and producer-sided fairness is an important requirement for a healthy marketplace. 
To capture fairness for multisides, \citet{burke2018balanced} introduce the task of two-sided fairness and employ the sparse linear method (SLIM)~\cite{NingK11SLIM} to address it.
~\new{\citet{PatroBGGC20FairRec} borrow notions from fair division~\cite{FairDivision} to model the two-sided fairness in recommendation. 
Specifically, they ensure the envy-freeness-up-to-one on the consumer side and maximin share guarantee of exposure on the producer side~\cite{Budish10EF1}.
Both fairness is treated at an individual level. 
~\citet{Wu0XT21TFROM} follow a similar way to model the individual fairness on both sides, where they ensure each individual consumer obtains equal satisfaction and each individual producer obtains equal (or proportional) exposure.
Both of the above two works adopt algorithms similar to Round-robin scheduling~\cite{OperatingSystem, round-robin1, round-robin2} to achieve the fairness in recommendation.}
\citet{suhr2019TwosidedShare} experiment with two-sided fairness in the context of ride-hailing platforms, where the two-sided objective is a linear interpolation of consumer and producer fairness metrics, like other works.


\subsection{Approaches to Achieving Fairness}
Motivated by the idea of constructing multiple objectives in recommendation \cite{DBLP:conf/recsys/JamborW10, DBLP:conf/chi/McNeeRK06}, most works on fairness in recommendation and ranking scenarios model the fairness as an extra loss.
It works as a supplement to the accuracy (quality) loss in the whole objective function \cite{10.1145/3109859.3109887, singh:fair-pg-rank, DBLP:conf/kdd/SinghJ18}, followed by employing the scalarization technique.
It is expected to achieve a Pareto optimal recommendation \cite{10.1145/2365952.2365962, 10.1145/2629350} when multiple objectives are concerned.
However, existing studies mostly depend on manually assigning weights for scalarization, whose Pareto optimality cannot be guaranteed. 

Recent studies have proposed to use the adversarial learning and causal graph reasoning techniques to achieve fairness in recommendation systems. For instance, \citet{Beigi_2020} propose an adversarial learning-based recommendation model with attribute protection, which can protect consumers from the private-attribute inference attack while simultaneously recommending relevant products to consumers. \citet{ijcai2019-456} find that the bias in recommendation is caused by unfair graph embeddings and thus propose a novel fairness-aware graph embedding algorithm $Fairwalk$ to achieve the statistical parity. \citet{bose2019compositional} combine the adversarial training with the graph representation learning together to protect sensitive features of consumers. They introduce an adversarial framework to enforce
fairness on graph embeddings. 
\new{Similarly, ~\citet{WuCSHWW21FairGO} propose a graph based adversarial learning method, \textit{FairGO}, to filter any sensitive information hidden in the data representation, where the fairness requirement is defined as not to expose sensitive features during the user modelling.}
The benefits of these algorithms lie in explicitly modelling the fairness into the representation embeddings; however, the models are based on more advanced techniques and they do not consider multisided fairness explicitly.

Fair Learning-to-Rank (LTR) is another popular research direction for achieving fairness in the community nowadays, and several recent works have raised the question of group fairness in rankings. Zehlike et al.~\cite{ZehlikeB0HMB17_FAIR} formulate the problem as a ``Fair top-$K$ ranking'' that aims to guarantee the occurrences of items within the protected group is above a minimum threshold in every prefix of the top-$K$ ranking list based on some pre-defined proportion.
Celis et al.~\cite{DBLP:conf/icalp/CelisSV18} propose a constrained maximum weight matching algorithm for ranking a set of items efficiently under a fairness constraint indicating the maximum number of items with each sensitive attribute being allowed in the top positions. Most recently, some works break the parity constraints restricting the fraction of items with each attribute in the ranking but extend the LTR methods to a large class of possible fairness constraints. For instance, Biega et al.~\cite{asia:equity-of-attention} aim to achieve amortized fairness of attention by making exposure proportional to relevance through integer linear programming. Singh et al.~\cite{DBLP:conf/kdd/SinghJ18} propose a more general framework that can achieve both individual fairness and group fairness solutions via a standard linear program and the Birkhoff-von Neumann decomposition~\cite{birkhoff1967lattice}.

\subsection{Recommendation with Multiple Objectives}
The studies on multi-objective optimization are rich and various approaches have been proposed \cite{moop}.
One significant characteristic of multi-objective optimization is that, usually, there does not exist a solution that satisfies all the objectives simultaneously.

Some studies have considered multiple objectives in personalized recommendation tasks \cite{DBLP:conf/recsys/JamborW10, 10.1145/2365952.2365962}. For instance, \citet{10.1145/2365952.2365962} construct multiple objectives including accuracy, diversity, and novelty.
And then a Pareto frontier is found to satisfy the mentioned objectives. However, manual scalarization (grid search) is still required. Besides, to the best of our knowledge, there are few studies on optimizing multiple objectives in group recommendation and we are among the first to address the multi-stakeholder fairness problem in recommendation from the multi-objective optimization perspective.
\vspace{3mm}

\new{\noindent\textbf{Novelty and Difference to Prior Works.}
Our study expands on prior works by studying interpolation-free optimization of multi-stakeholder fairness problems.
Compared to prior works, we highlight the novelty and difference of our work as follows. 
Firstly, we are the first to propose a general framework for multi-stakeholder fairness-aware recommendation with theoretical guarantee that the final solution is Pareto stationary (Pareto optimal under mild assumptions). 
Prior works hardly satisfy the Pareto optimality. 
Secondly, we propose a way to differentiate the fairness metrics on both the consumer side and the producer side, so that we can directly optimize these fairness constraints during the model training in an end-to-end way. 
However, most metrics defined in prior works are not differentiable. 
Thus, they can only use a post-processing method to audit the recommendation list after obtaining the relevance scores between users and items. 
Thirdly, we employ the Frank-Wolfe Solver and propose a method to learn the scaling factors on multiple objectives adaptively during training. 
However, most prior works require additional scaling factor tuning. 
In addition, we utilize the stochastic ranking policy, which is capable of distributing the exposure among producers more equitably, whereas most prior works adopt the static ranking.}

\section{Preliminaries}

\subsection{List of Notations}
The notations we used in this paper are shown in Table~\ref{table:notation1}. 

\begin{table}[h]
\centering

\caption{\label{tab:formulation}Description of Notations.}
\vspace{1mm}
\resizebox{1.0\textwidth}{!}{
\begin{tabular}{cl}
\toprule
    Notations & Description  \\
\midrule
$\mathcal{U}, \mathcal{I}$ & The set of users and items\\ 
$\mathcal{D}_u$ & The set of training triplets\\
$\mathcal{L}_i$ & The $i^{th}$ objective\\
$\Theta$ & Embeddings for users and items\\
$\alpha$ & The scaling factor on each objective\\
$b, K$ & Batch size and the maximum length of the recommendation list\\
$n^C, n^P$ & Number of groups on the consumer side and the producer side\\
$t$ & The total number of objectives in the model\\
$m, n$ & The number of fairness constraints on the consumer side and the producer side\\
$l_i$ & The relevance score between the $i^{th}$ item and the current query (user)\\
$r_i$ & The rank of the $i^{th}$ item w.r.t. the current query\\
$\tilde{r}_i$ & The smooth rank of the $i^{th}$ item w.r.t. the current query under the stochastic policy\\
$\gamma$ & User's patience factor that controls the depth of browsing a list of items\\
$\tau$ & Temperature that controls the smoothness of ranks under the stochastic policy\\

$\bm{Y}\in\mathbb{R}^{|\mathcal{U}|\times|\mathcal{I}|}$ & The binary relevance matrix between items and users\\
$\bm{s}^{G_i}\in\mathbb{R}^K$ & The satisfaction representation for group $G_i$ on the consumer side\\
$\bm{m}^{G_i}\in\mathbb{R}^b$ & The mask of group $G_i$ on the consumer side\\
$\bm{N}\in\mathbb{R}^{b\times K}$ & The matrix of NDCG@1 to NDCG@$K$ for all the users in a batch\\
$\epsilon, \epsilon^*\in\mathbb{R}^{n_p}$ & System exposure and target exposure on the producer side\\
$\bm{R}\in\mathbb{R}^{b\times|\mathcal{I}|}$ & Ranks for all items with respect to a batch of users\\
$\bm{E}, \bm{E}^*\in\mathbb{R}^{b\times|\mathcal{I}|}$ & System and target exposure matrix of all items w.r.t. a batch of users\\
$\bm{c}\in\mathbb{R}^{|\mathcal{I}|}$ & Group label of all items\\
\bottomrule
\end{tabular}}
\label{table:notation1}
\end{table}

\subsection{Problem Formulation}
\label{sec:problem_task}
We consider a top-$K$ item recommendation task in this paper which takes the user implicit feedback as input. We denote the set of all users and items as $\mathcal{U}$ and $\mathcal{I}$, respectively. For each user $ u $, the user preference data is represented by a set of items he/she has interacted with as $\mathcal{I}_{u}^{+}:=\{i \in \mathcal{I}|\bm{Y}_{u,i}=1\}$ where $\bm{Y} \in \mathbb{R}^{\mathcal{|U|} \times \mathcal{|I|}}$ is the binary implicit feedback rating matrix. We then split $\mathcal{I}_{u}^{+}$ into a training set $\mathcal{S}_{u}^{+}$ and a test set $\mathcal{T}_{u}^{+}$, requiring that  $ \mathcal{S}_{u}^{+} \cup \mathcal{T}_{u}^{+} = \mathcal{I}_{u}^{+} $ and $ \mathcal{S}_{u}^{+} \cap \mathcal{T}_{u}^{+} = \emptyset$. Then the top-$K$ recommendation is formulated as: given the training item set $ \mathcal{S}_{u}^{+} $, and the non-empty test item set $ \mathcal{T}_{u}^{+}$ of user $ u $, the model aims to recommend an ordered set of $K$ items $ \mathcal{X}_{u} $ such that $ |\mathcal{X}_{u}| = K $ and $ \mathcal{X}_{u} \cap \mathcal{S}_{u}^{+} = \emptyset $.

As aforementioned, our goal of this work is to well balance all of the following objectives on the quality and fairness in an end-to-end recommendation framework. We employ the notion of \textit{group fairness}~\cite{DworkHPRZ12FairnessThroughAwareness} on both the consumer side and the producer side, which suggests treating different groups in a fair way.
\begin{enumerate}
    \item \textit{Recommendation Accuracy}: This is to ensure that the recommendation model is capable of capturing consumers' real preference. The recommendation quality can be evaluated by a matching score between $\mathcal{T}_{u}^{+}$ and $\mathcal{X}_{u}$, such as Recall@$K$ or NDCG@$K$ (Normalized Discounted Cumulative Gain). 
    \item \textit{Consumer-sided Fairness}: This is to guarantee that the consumers belonging to different demographic groups receive the same level of satisfaction. Specifically, we aim to minimize the group-level satisfaction difference between any two groups of consumers. The definition will be detailed in Section~\ref{sec:objective}.
    \item \textit{Producer-sided Fairness}: This is to avoid the producers belonging to minority groups receive an extremely high/low opportunity of being exposed. Specifically, we aim to minimize the difference between the current computed exposure (system exposure) and the ideal exposure (target exposure) of producers. The definition will be detailed in Section~\ref{sec:objective}.
\end{enumerate}

\section{Objective Construction}
\label{sec:objective}
In this section, we demonstrate the formulation of each objective: recommendation quality, consumer-sided fairness, and producer-sided fairness, followed by offering the technical details for making all of these objectives differentiable for an end-to-end training
~\footnote{\new{In this paper, we use \textit{producer-sided fairness} to represent the fairness on the item side, although the producer information is not provided in most datasets.
There are two reasons for choosing this term: (1) We would like to make the term consistent with prior works~\cite{Wu0XT21TFROM, suhr2019TwosidedShare,MehrotraMBL018CounterTradeoff} in community; (2) People are building systems for humans not for items: achieving fairness on items actually benefits the producers.
We will collect the supplier/producer information of items in future work.}
Additionally, \textit{user} and \textit{consumer} share the same meaning throughout the remainder of this paper.}.

\subsection{Objective for Recommendation Quality}
In order to ensure the recommendation quality, we adopt the Bayesian Pairwise Ranking (BPR) model proposed by~\citet{rendle2012bpr}. Denoting those items that are unobserved by user $u$ in $\mathcal{S}_{u}^{+}$ as $\mathcal{S}_{u}^{-}:=\mathcal{I} \setminus \mathcal{S}_{u}^{+}$, we define a new training set in which each component is a triplet:
\begin{equation}
\mathcal{D}_{u}:=\left\{(u, i^+, i^-) | i^+ \in \mathcal{S}_{u}^{+} \wedge i^- \in \mathcal{S}_{u}^{-}\right\} \,.
\end{equation}
Then the goal of the recommendation model is to generate a total ranking $>_u$ of all items for each user $u$. The binary relation $>_u$ is required to be a total order on the set of items $\mathcal{I}$. The relation $i^+ >_u i^-$ specifies that user $u$ prefers item $i^+$ over item $i^-$. Thereby, we aim to maximize:
\begin{equation}
p\left(\Theta |\{>_{u}\}_{\mathcal{D}_{u}}\right) \propto p\left(\{>_{u}\}_{\mathcal{D}_{u}} | \Theta\right) p(\Theta)\,,
\end{equation}
where $\Theta=\llbracket\Theta^U, \Theta^I\rrbracket$ is the model parameter containing the user and item embeddings, and $\{>_{u}\}_{\mathcal{D}_{u}}$
denotes the observed preferences in the training data. 
We aim to identify
the parameters $\Theta$ that maximize this posterior over all users
and all pairs of items. Assuming that users act independently, we have:
\begin{equation}
    p\left(\{>_{u}\}_{\mathcal{D}_{u}} |  \Theta \right)=\prod_{(u, i^+, i^-) \in \mathcal{D}_{u}} p\left(i^+>_{u} i^- | \Theta\right).
\end{equation}
We define the probability that a user prefers item $i^+$ over item $i^-$ as:
\begin{align}
p\left(i^+>_{u} i^- | \Theta
  \right) =\sigma\left(\hat{x}_{ui^+}(\Theta)-\hat{x}_{ui^-}(\Theta)\right),
\label{eq:rankprob}
\end{align}
where $\hat{x}_{ui}(\Theta) = \langle \Theta^U_u, \Theta^I_i \rangle \,$, denoting the inner product between two embeddings.
If we adopt a normal distribution as the prior for
$p(\Theta)$, then we can formulate the optimization
  objective as:
\begin{align}
\mathcal{L}^{Accuracy}&=\argmax\limits_{\Theta}\ln p\left(\Theta |\{>_{u}\}_{\mathcal{D}_{u}}\right) \nonumber\\
&=\argmin\limits_{\Theta}\sum_{(u, i^+, i^-) \in \mathcal{D}_{u}} -\ln \sigma\left(\hat{x}_{ui^+}(\Theta)-\hat{x}_{ui^-}(\Theta)\right)+\lambda_{\Theta}||\Theta||_2^{2}.
\label{eq:BPROPT}
\end{align}
We can optimize this via stochastic gradient descent by repeatedly
drawing triples $(u,i^+,i^-)$ randomly from the training set and updating
the model parameter $\Theta$.

\subsection{Fairness Objectives for Multi-stakeholder}
Fairness attracts more attention in current information retrieval systems, which has a huge impact on the multi-stakeholder marketplaces. In our proposed framework, we resort to group fairness on both the consumer (user) side and the producer (item) side. We define four fairness constraints (two for each side) with respect to different attributes in our model which are summarized in Table~\ref{table:fairness_constraint}.

\begin{table}[t]
\centering
\caption{\label{tab:formulation}Four attribute-based fairness constraints on the consumer side and the producer side.}
\vspace{1mm}
\begin{tabular}{lll}
\toprule
Stakeholder & \multicolumn{2}{c}{Attribute-based Fairness Constraints}\\
\midrule
Consumer & \textbf{Gender}-based & \textbf{Age}-based\\
Producer & \textbf{Popularity}-based & \textbf{Genre}-based\\
\bottomrule
\end{tabular}
\label{table:fairness_constraint}
\end{table}

\subsubsection{Consumer Fairness Constraint}
\label{subsub:consumer-sided fairness}
It has been shown that sensitive features affect the satisfaction of consumers in recommendation~\cite{DBLP:conf/cikm/ZhuHC18}. 
\new{For instance, \citet{ekstrand2018all} and \citet{neophytou2022revisiting} demonstrate that recommendation performance can vary across demographic groups and \citet{mehrotra:demographics-of-search} report similar observations in the context of web search. For addressing this, the fairness of consumers is generally defined as the fairness in quality of service, such as ensuring those consumers belonging to different demographic groups (with different sensitive features) experience comparable recommendation quality.
}
In this work, we follow the similar line to model the group fairness on the consumer side.
Specifically, we aim to make the satisfaction of different groups with sensitive features be ideally equal. 
Here we adopt the NDCG@$K$, a widely-used ranking metric, to measure the satisfaction of consumers. 

However, since items ranked at higher positions generally receive more attention from consumers, we argue that the fairness at each prefix of the ranking is also significant. As a result, solely considering the recommendation equality for an entire ranking list is not enough. Thereby, we construct the $\textbf{s}^{G_i}\in \mathbb{R}^{K}$ as a satisfaction vector for the group $G_i$ among the consumers, where the $k^{th}$ entry of $\textbf{s}^{G_i}$ equals to $\text{NDCG@}k$, $k=1, ..., K$. Assuming there are $n^C$ groups among the consumer side, the fairness constraint on a group level can be defined as the mean of the pair-wise satisfaction difference across all groups:
\begin{equation}
    \mathcal{L}^{C-Fair}=\frac{1}{\tbinom{n^C}{2}}\sum_{1\leq i<j\leq n^C}\left\|\textbf{s}^{G_i} -\textbf{s}^{G_j}\right\|^2_2\,,
\label{eq:fairness_user}
\end{equation}
where $\tbinom{n^C}{2}$ is the number of combinations between different groups. 

More specifically, considering a batch of $b$ users (consumers), we define the satisfaction of group $G_i$ as:
\begin{equation}
    \textbf{s}^{G_i} = \frac{ \textbf{m}^{G_i}\cdot\textbf{N}}{||\textbf{m}^{G_i}||_1} \,,
\label{eq:user_satisfaction}
\end{equation}
where $\textbf{m}^{G_i}\in\mathbb{R}^b$ is a multi-hot vector, $\textbf{N}\in \mathbb{R}^{b\times K}$ is a matrix containing NDCG@$1$ to NDCG@$K$ for all users in a batch. The method of computing the NDCG values in a differentiable way is detailed in section~\ref{section:smooth_method}. The denominator is used to compute the number of users in the group $G_i$.

It is worthy to notice that any kinds of attributes can be adopted for defining the mask $\textbf{m}^{G_i}$ in Eq.~\ref{eq:user_satisfaction}. In this work, we focus on two types of disparities regarding the two most common and sensitive attributes on the consumer side.

\vspace{1mm}
\textbf{Gender-based Fairness}. Gender is one of the most sensitive attributes of humans and many works have already presented insightful observations and analysis on gender bias in Internet services~\cite{DBLP:conf/chi/KayMM15, google_ceo}. Thus, we construct the gender mask $\textbf{m}^{G_{1}}$ and $\textbf{m}^{G_{2}}$ for females and males and aim to minimize the satisfaction difference between these two groups.
\textit{\textcolor{red}{Note that gender is treated as a binary class due to the available labels in the datasets. 
We do not intend to suggest that gender identities are binary, nor support any such assertions.}}

\vspace{1mm}
\textbf{Age-based Fairness}. Other than gender, we also consider fairness with respect to the age attribute. We split the age into 7 stages (0-17, 18-24, 25-34, 35-44, 45-49, 50-55, 56+) following the criterion in the MovieLens datasets~\cite{DBLP:journals/tiis/HarperK16} and construct $\textbf{m}^{G_{a_i}}$ as the mask for the $i$-th age group. Then optimizing the age-based fairness is to minimize the difference of satisfaction for all age groups, as described in Eq.~\ref{eq:fairness_user}.

\subsubsection{Producer Fairness Constraint}
Previous works in recommendation mainly assume that the users are the only stakeholder in a recommendation system; however, the items should also be taken into consideration since they represent the benefits of the producers~\cite{PatroBGGC20FairRec, suhr2019TwosidedShare}, which are an equally significant stakeholder in a commerce marketplace. Thus, we model the producer-sided fairness to guarantee the satisfaction of producers. 

In comparison to consumers, producers are more concerned with profits in these marketplaces, which are highly related to the exposure of their products/items. Unfair exposure distribution on items may make certain producers unsatisfied and hence leave the platform. This may reduce the market's diversity, which in turn may harm the consumers' utility. Therefore, for the group fairness on producers, the goal is to find a ranking strategy that can offer a fair probability of exposure on items based on their merits. However, one of the key challenges, as mentioned in~\cite{DBLP:conf/cikm/DiazMEBC20}, is that a single fixed ranking for a query (in retrieval) or user (in recommendation) tends to limit the ability of an algorithm to distribute exposure amongst relevant items. For a static ranking, (\romannumeral1) some relevant items may receive more exposure than other relevant items, and (\romannumeral2) some irrelevant items may receive more exposure than other irrelevant items. Therefore, we aim to find a policy that samples a permutation from a distribution over the set of all permutations of $|\mathcal{I}|$ items, and such a stochastic ranking policy should be able to force all items to receive a fair exposure proportional to their merits, thus achieving fairness with respect to the exposure of items in expectation. 

Assuming there are $n^P$ groups among all items, the fairness constraint on the producer side can be defined as the difference between two exposure vectors:
\begin{equation}
    \mathcal{L}^{P-Fair}=\left\|\epsilon - \epsilon^*\right\|^2_2,
\label{eq:fairness_item}
\end{equation}
where $\epsilon\in \mathbb{R}^{n^P}$ is a vector representing the distribution of exposure on items from different groups,  $\epsilon^*\in \mathbb{R}^{n^P}$ is the ideal exposure distribution proportional to the true relevance of items. We refer to the $\epsilon$ and $\epsilon^*$ as the \textit{system exposure} and \textit{target exposure} of the system, respectively. Hereafter, we demonstrate how to model $\epsilon$ and $\epsilon^*$.



For computing the system exposure $\epsilon$, we need to first rank all the $|\mathcal{I}|$ items for $b$ users in a batch based on the relevance score. Under a static ranking policy, this is generally achieved by sorting the items in a descending order based on the predicted preference score between users and items.  We can thus obtain a matrix $\bm{R}\in\mathbb{R}^{b\times |\mathcal{I}|}$ containing the ranks of all items for $b$ queries (users). (As noted before, we are interested in a stochastic policy instead. The method for obtaining the matrix $\bm{R}$ under a stochastic policy will be detailed in section~\ref{section:smooth_method}.) To compute the exposure, we adopt a well-known user browsing model, the position-based model~\cite{10.1145/1416950.1416952, DBLP:conf/cikm/DiazMEBC20}, that assumes a user's probability of visiting a position decreases exponentially with the rank. Then we compute the exposure of all items in a batch as $\bm{E} = \gamma^{\bm{R}}\in\mathbb{R}^{b\times |\mathcal{I}|}$, where $\gamma$ represents the patience factor that controls how deep a user is likely to browse in a ranked item list. 

We denote the group label on all items in $\mathcal{I}$ sorted by the item id as $\bm{c}\in\mathbb{R}^{ |\mathcal{I}|}$, where each entry contains the group label that the corresponding item belongs to. However, the items are displayed in different orders for each user, we thus denote the group label of items ranked for the $i^{th}$ user based on the ranking order as $\bm{c}^i\in\mathbb{R}^{ |\mathcal{I}|}$, which is a permutation of $\bm{c}$. Then, the $\epsilon$ is computed as:
\begin{equation}
    \epsilon_k=\frac{\sum\limits_{i=1}^{b}\sum\limits_{\mathbbm{1}_{{\bm{c}^i_j=k}}}{\bm{E}_{ij}}}{b\cdot||\mathbbm{1}_{{\bm{c}_j=k}}||_1}.
\label{eq:sys_exposure}
\end{equation}
Here the $\mathbbm{1}_{\bm{c}_j=k}$ is the indicator vector, where a ``1'' at position $j$ refers to $\bm{c}_j=k$ and a ``0'' otherwise. The $k^{th}$ entry of $\epsilon$ contains the average exposure on all items belonging to the $k^{th}$ group.  

As for the target exposure $\epsilon^*$, we assume all relevant items should have the same high probability of being selected to the top of the ranking, while other irrelevant items should equally share the rest amount of exposure at a low level. Given the binary relevance label in the training set as $\bm{Y}\in\mathbb{R}^{|\mathcal{U}|\times|\mathcal{I}|}$, we assume the number of relevant items for each query (user) is $t_i$, which is obtained through counting the number of ones in each row of $\bm{Y}$. Then the target exposure of all items in a batch can be computed as:
\begin{equation}
    \bm{E}^*_{ij}=\left\{
    \begin{aligned}
    &\frac{1}{t_i}\sum_{m\in[1, t_i]}\gamma^m,\qquad\qquad\text{if } \bm{Y}_{ij}=1,\\
    &\frac{1}{|\mathcal{I}|-t_i}\sum_{m\in[t_i+1, |\mathcal{I}|]}\gamma^m, \quad\text{otherwise.}
    \end{aligned}\right.
\end{equation}

Thus, we can construct the target exposure in a similar way as Eq.~\ref{eq:sys_exposure}:
\begin{equation}
    \epsilon^*_k=\frac{\sum\limits_{i=1}^{b}\sum\limits_{\mathbbm{1}_{{\bm{c}^i_j=k}}}{\bm{E}^*_{ij}}}{b\cdot||\mathbbm{1}_{{\bm{c}_j=k}}||_1}.
\label{eq:tgt_exposure}
\end{equation}

Then we can use Eq.~\ref{eq:fairness_item}, Eq.~\ref{eq:sys_exposure}, and Eq.~\ref{eq:tgt_exposure} to optimize for the producer-sided fairness.

\vspace{1mm}
We consider two types of group fairness constraints on the producer side. Both are based on the same framework defined in Eq.~\ref{eq:fairness_item}, while the only difference lies in the different construction of the group label $\bm{c}$ during the modelling. 

\vspace{1mm}
\textbf{Popularity-based Fairness}. The ``superstar economics'' \cite{MehrotraMBL018CounterTradeoff, baranchuk2011economics} always occurs in real-world recommendation scenarios, where a small number of most popular artists/items/products possess most of the attention of consumers. A major side-effect of the ``superstar economics'' is the impedance to producers on the tail-end of the spectrum, who struggle to attract consumers and are not satisfied with the marketplace.

To construct the popularity-based group label, we rank all the $|\mathcal{I}|$ items based on their occurrences in the dataset from the highest to the lowest and evenly split them into 5 groups labelled from $4$ to $0$, where each group contains $20\%$ of items. 

\vspace{1mm}
\textbf{Genre-based Fairness}. We do not expect any specific genre of items to receive too much or too little exposure in a marketplace; therefore, genre-based fairness is also worthy of taking into consideration. We only use the MovieLens datasets to investigate this fairness and make use of all the $18$ movie genres in the datasets \cite{DBLP:journals/tiis/HarperK16}. 

\subsection{Differentiable Approximation of the Ranking}
\label{section:smooth_method}

In our formulation, relevance is defined as a function of item rankings, but the sorting operation is inherently non-differentiable. To mitigate this problem, we adopt the continuous approximation of the ranking function proposed in~\citep{DBLP:conf/cikm/WuCZZ09,DBLP:journals/ir/QinLL10} that is amenable to gradient-based optimization.
The key insight behind these approximations lies in defining the rank of an item in terms of the pairwise preference with every other item in the collection:
\begin{align}
    r_i = 0.5 + \sum_j^{n}{\sigma'(l_j - l_i)}, \,
    \text{where }\; \sigma'(x) =
                        \begin{cases}
                            1, & \text{if } x > 0 \\
                            0.5, & \text{if } x = 0 \\
                            0, & \text{if } x < 0
                        \end{cases}.
\end{align}
Here the $l_i$ is the relevance score between the $i^{th}$ item to the query (user), which is computed through the inner-product between the user embedding and the item embedding. The discrete function $\sigma'(\cdot)$ is typically approximated using the differentiable sigmoid function.

Given the approximated differentiable ranks of items, it is then straightforward to derive an optimization objective for standard relevance metrics---\eg, discounted cumulative gain (DCG)---that can be directly optimized using the gradient descent. Assuming that we consider the items in a ranking up to position $K$, then the SmoothDCG is defined as:
\begin{align}
    \text{SmoothDCG}= \sum_{i=1}^{K}{\frac{l_i}{\text{log}_2(r_i + 1)}} \,.
\end{align}
Therefore, we adopt the normalized version of such SmoothDCG when constructing the fairness objective in Eq.~\ref{eq:user_satisfaction} on the consumer side during training, but still, use the original NDCG in the evaluation phase.



As for the producer side, in order to mitigate the similar non-differentiable operation when constructing the ranking position matrix $\bm{R}$ and also adopt a stochastic ranking policy rather than a static one, we use the Plackett-Luce model~\cite{Plackett1975TheAO} for constructing a ranking by sampling items sequentially, followed by adopting the Gumbel Softmax technique proposed in \cite{maddison2016concrete, DBLP:conf/wsdm/BruchHBN20}. For a single query (user), we recall that the sampling probability of an item by using a static Plackett-Luce policy~\cite{Plackett1975TheAO} is:
\begin{equation}
    p_i = \frac{\exp(l_i)}{\sum_{j\in \mathcal{I}}\exp(l_j)} \,.
\label{PL_policy}
\end{equation}
As aforementioned, the static policy will limit the ability of the ranking algorithm for fairly distributing the exposure. 
Thus, what the stochastic ranking policy specifically performs here is: (\romannumeral1) reparameterizing the probability distribution by adding independently-drawn noise $\zeta$ sampled from the Gumbel distribution to $l$ and (\romannumeral2) sorting items by the ``noisy'' probability distribution $\tilde{p}_i$:
\begin{equation}
    \tilde{p}_i = \frac{\exp(l_i+\zeta_i)}{\sum_{j\in \mathcal{I}}\exp(l_j+\zeta_j)} \,.
\end{equation}
However, the sorting operation itself is non-differentiable either.
To address this, we instead compute the smooth rank~\cite{DBLP:conf/cikm/WuCZZ09} for each item in the ranking as follows:
\begin{equation}
    \tilde{r}_i = \sum\limits_{j\in \mathcal{I}, j\neq i}\left(1+\exp\left(\frac{\tilde{p}_i-\tilde{p}_j}{\tau}\right)\right)^{-1} \,,
\end{equation}
where the temperature $\tau$ is a hyperparameter that controls the smoothness of the approximated ranks. Then the exposure $\bm{E}$ in Eq.~\ref{eq:sys_exposure} is computed based on this smooth rank. 

We now achieve a differentiable way for modelling both the consumer-sided fairness and the producer-sided fairness for an end-to-end training.



\section{Multi-FR: Multi-stakeholder Fairness-aware Recommendation}
\label{sec:multifr}

In this section, we introduce our proposed framework, \textit{Multi-FR}, for fairness-aware recommendation in multi-stakeholder marketplaces. In conventional recommendation systems, the main aim lies in satisfying consumers. However, it has been shown in recent studies~\cite{PatroBGGC20FairRec} that solely optimizing the satisfaction of consumers may jeopardize the benefits of producers who are essential participants in two-sided markets. Thus, to achieve a personalized, satisfactory, and fair recommendation simultaneously in one joint framework is significant in both academia and industry.

Traditionally, these aspects are modelled as specific objectives and combined by summation with manually set scaling factors. However, utilizing hand-crafted factors has two major drawbacks. First, these scaling factors incur tedious hyper-parameter tuning. This would cost many trials and substantial computation resources to identify appropriate scaling factors, especially when the number of objectives is huge. Second, each objective in the summed objective function may need a different magnitude of scaling values in the training process. Setting one fixed value is not capable of dynamically balancing all of these objectives well during the training process.


To tackle the aforementioned problems, we treat the fairness-aware recommendation as a multi-objective optimization problem and propose a framework for optimizing multiple objectives jointly. Before diving into the final framework, we start from providing some major techniques and theoretical guarantees in the multi-objective optimization in order to better illustrate the entire picture.

\subsection{Multi-Objective Optimization}
A multi-objective optimization task is usually defined as optimizing a set of possibly conflicting objectives. Given a set of objectives, the MOO aims to find a solution that can optimize all objectives simultaneously:
\begin{equation}
\resizebox{0.6\textwidth}{!}{
\begin{math}
    \min\limits_{\theta}\mathcal{L}(\theta)=\min\limits_{\substack{\theta^c\\ \theta^{s_1}, ...,  \theta^{s_t}}}\mathcal{L}(\theta^c, \theta^{s_1}, ..., \theta^{s_t})
    =\min\limits_{\substack{\theta^c\\ \theta^{s_1}, ...,  \theta^{s_t}}}
    \left[
    \begin{array}{c}
          \mathcal{L}_1(\theta^c, \theta^{s_1}) \\
          \mathcal{L}_2(\theta^c, \theta^{s_2}) \\
          \vdots \\
          \mathcal{L}_t(\theta^c, \theta^{s_t}) \\
    \end{array}
    \right]^\intercal,
\end{math}}
\end{equation}
where $\mathcal{L}$ is the full objective, $\mathcal{L}_1$, ..., $\mathcal{L}_t$ are $t$ different objectives, respectively. $\theta^c$ is the common parameter shared by all objectives, while $\theta^{s_1}$, ..., $\theta^{s_t}$ are the objective-specific parameters.

Notice that one of the key characteristics of an MOO problem is that a solution that can optimize each objective to an ideal situation may not exist. This is exactly due to the conflict and correlation among the objectives as discussed before. The optimal solution of an MOO problem should balance all the objectives, which is called the Pareto optimality.

\vspace{3mm}

\noindent \textbf{Definition 1. Pareto Optimality}
\indent\begin{enumerate}
    \item A solution $\theta_1$ \textbf{dominates} another solution $\theta_2$ if for all objectives $\mathcal{L}_i(\theta_1^c, \theta_1^{s_i})\leq \mathcal{L}_i(\theta_2^c, \theta_2^{s_i})$, where $i\in\{1,..., t\}$, and there exists at least one objective $j\in\{1,..., t\}$, where\\ $\mathcal{L}_j(\theta_1^c, \theta_1^{s_j})< \mathcal{L}_j(\theta_2^c, \theta_2^{s_j})$.
    \item A solution is of \textbf{Pareto optimality} if there does not exist any other solution that dominates it. In this case, we also call the solution is \textbf{Pareto optimal}.
    \item There is usually more than one solution being Pareto optimal in an MOO problem. The set of such solutions is called \textbf{Pareto set}, which is the solution set of an MOO problem. The curve of the points in the Pareto set is called the \textbf{Pareto frontier}.
\end{enumerate}

\subsection{Multiple Gradient Descent Algorithm}
The multiple gradient descent algorithm (MGDA)~\cite{desideri2012multiple} is one of the most effective methods for MOO. It can reach an optimal point for all objectives with theoretical guarantee. Borrowing the idea from the gradient descent on a single objective, the MGDA can be regarded as an extension of the gradient-based algorithm on multiple objectives. The overall objective of solving an MOO problem by MGDA is usually a weighted summation of $t$ single objectives, defined as:
\setlength{\textfloatsep}{0.1pt}
\begin{equation}
      \mathcal{L}(\theta)=\mathcal{L}(\theta^c, \theta^{s_1}, ..., \theta^{s_t})= \sum\limits_{i=1}^t\alpha_i\cdot\mathcal{L}_i(\theta^c, \theta^{s_i}) \,,
\end{equation}
\vspace{-0.1mm}
where the coefficients of all the objectives satisfy $\sum\limits_{i=1}^t \alpha_i=1$ and $\alpha_i\geq 0$, for $i=1,...,t$.

It is hard to find direct conditions for Pareto optimality.
Therefore, we introduce the Pareto stationarity, which is a necessary condition for Pareto optimality in an MOO problem~\cite{Condition_Pareto1, Condition_Pareto2, desideri2012multiple}.
A Pareto optimal solution must be Pareto stationary, while the reverse may not hold.

\vspace{3mm}
\noindent \textbf{Definition 2. Pareto Stationarity}\\
A solution $\theta^*$ is of \textbf{Pareto stationarity} if it satisfies all of the following conditions:
\begin{enumerate}
    \item $\sum\limits_{i=1}^t \alpha_i=1$, $\alpha_i\geq 0$, for $i=1,...,t$,
    \item $\sum\limits_{i=1}^t\alpha_i\nabla_{\theta^{*c}}\mathcal{L}_i(\theta^{*c}, \theta^{*s_i})=0$,
    \item $\nabla_{\theta^{*s_i}}\mathcal{L}_i(\theta^{*c}, \theta^{*s_i})=0$, for $i=1,...,t$.
\end{enumerate}
\vspace{2mm}

The above conditions are also known as the \textbf{Karush}-\textbf{Kuhn}-\textbf{Tucker} \textbf{(KKT)} \textbf{conditions}, which was first proposed by \citet{kuhn1951}.

\vspace{1mm}

The MGDA leverages the KKT conditions to solve the MOO problem, which are necessary for optimality. \citet{SenerK18MTLasMOO} propose to solve a quadratic-form constrained minimization problem defined as follows:
\begin{equation}
\setlength\abovedisplayskip{2pt}
\setlength\belowdisplayskip{2pt}
\begin{aligned}
    \min\limits_{\alpha_1, \alpha_2, ..., \alpha_t} & \left\| \sum\limits_{i=1}^t\alpha_i\cdot\nabla_{\theta^{c}}\mathcal{L}_i(\theta^c, \theta^{s_i})\right\|_2^2 \,,\\
    \text{s.t.}, & \sum\limits_{i=1}^t\alpha_i=1, \alpha_i\geq0, \text{for } i=1,...,t \,.
\end{aligned}
\label{eq:quad_opt}
\end{equation}

Given Eq.~\ref{eq:quad_opt}, there are two situations for the final solution: the final solution is Pareto stationary if the solution to this optimization problem makes the Euclidean norm equals to $0$; otherwise, the solution offers a common descent direction which benefits all the objectives as proved by~\cite{desideri2012multiple}. Therefore, one can use the single-objective gradient descent for optimizing the objective-specific parameters $\theta^{s_i}$ on $t$ different objectives and employ the obtained solution to the above equations for updating the common parameters $\theta^{c}$.

\subsection{Solving the MOO Problem}

We first introduce a special case where there are only two objectives in the loss function:
\begin{equation}
    \min\limits_{\alpha\in[0,1]}\left\| \alpha\cdot\nabla_{\theta^{c}}\mathcal{L}_1(\theta^c, \theta^{s_1})+(1-\alpha)\cdot\nabla_{\theta^{c}}\mathcal{L}_2(\theta^c, \theta^{s_2})\right\|_2^2 \,.
\end{equation}

The analytical solution to this quadratic problem is:

\begin{equation}
\label{eq:twoMOO}
    \alpha^*=\frac{\left(\nabla_{\theta^{c}}\mathcal{L}_2(\theta^c, \theta^{s_2})-\nabla_{\theta^{c}}\mathcal{L}_1(\theta^c, \theta^{s_1})\right)^\intercal\nabla_{\theta^{c}}\mathcal{L}_2(\theta^c, \theta^{s_2})}{\left\| \nabla_{\theta^{c}}\mathcal{L}_1(\theta^c, \theta^{s_1})-\nabla_{\theta^{c}}\mathcal{L}_2(\theta^c, \theta^{s_2})\right\|_2^2},
\end{equation}
where the $\alpha^*$ should be clipped into $[0,1]$ as $\alpha^*=\max(\min(\alpha, 1), 0)$.

Although there are no analytical solutions for more than two objectives in an MOO problem, we can still borrow the analytical solution of two objectives to conduct the line search efficiently. 
\new{This technique was proposed by~\citet{DBLP:conf/icml/Jaggi13} to accelerate the convergence of the Frank-Wolfe algorithm~\cite{Frank-Wolf}. 
Specifically, as shown in Algorithm~\ref{alg: Frank-Wolfe Solver}, the procedure with the same idea of Eq.~\ref{eq:twoMOO} is regarded as a subroutine to compute the $w^*$ on line 6.}
The scaling factors generated by the Frank-Wolfe Solver satisfy the KKT conditions aforementioned.

\begin{algorithm}[t!]
\caption{Frank-Wolfe Solver \cite{Frank-Wolf, DBLP:conf/icml/Jaggi13, SenerK18MTLasMOO}}
\SetKwFunction{algo}{algo}\SetKwFunction{proc}{proc}
\label{alg: Frank-Wolfe Solver}
\SetAlgoLined
\KwIn{
 $t \leftarrow$ number of objectives\\
 \qquad\quad$\theta \leftarrow$ model parameters: $(\theta^c, \theta^{s_1}, ...,  \theta^{s_t}) $\\}
\KwOut{A list of learned scaling coefficients: $\alpha_1$, ..., $\alpha_t$}

 \text{Random Initialization}: $\bm{\alpha}=(\alpha_1, ..., \alpha_t)$ \text{satisfying the constraints in Eq.~\ref{eq:quad_opt}}.\\ 
 \text{Precompute} $\bm{M}$, $\forall i,j \in\{1, ..., t\}:$ \\ \qquad $\bm{M}_{ij}=(\nabla_{\theta^{c}}\mathcal{L}_i(\theta^c, \theta^{s_i}))^\intercal(\nabla_{\theta^{c}}\mathcal{L}_j(\theta^c, \theta^{s_j}))$\\
\SetKwRepeat{REPEAT}{repeat}{until}
\REPEAT{$w^*$ \text{converge} or maximum iteration reaches}{
    $i^*=\argmin_r\sum_i\alpha_i\bm{M}_{ri}$\\
    $w^*=\argmin_w(w\bm{e}_{i^*}+(1-w)\bm{\alpha})^\intercal \bm{M} (w\bm{e}_{i^*}+(1-w)\bm{\alpha})$ $\qquad\leftarrow$\text{\new{Using Procedure 1}}\\
    $\bm{\alpha}=w^*\bm{e}_{i^*}+(1-w^*)\bm{\alpha} \quad(\bm{e}_{i^*} \text{ is the unit vector})$
    }
 \Return $\bm{\alpha}=(\alpha_1, ..., \alpha_t)${}\\
 
 \SetAlgoLined
 \nonl\new{\SetKwProg{myproc}{Procedure 1: Solving $\argmin_{w\in[0,1]}||wx_1+(1-w)x_2||^2_2$}{}{}
  \myproc{}{
  \setcounter{AlgoLine}{0}
  $w^*=\frac{(x_2-x_1)^\intercal x_2}{||x_1-x_2||^2_2}$\\
  $w^*=\max(\min(w^*, 1), 0)$\\
  \Return $w^*$\;}}
  
\end{algorithm}
\vspace{0.2mm}
\setlength{\textfloatsep}{2pt}

\subsection{The Multi-FR Framework and the Overall Objective}
Thereby, we propose the \textit{Multi-FR} framework to utilize the scaling factors that satisfy the KKT conditions to generate Pareto stationary (can be regarded as Pareto optimal under mild assumptions in real world) solutions which can smoothly balance the recommendation quality and multisided fairness. The algorithm is shown in Algorithm~\ref{alg: Multi-FR}. 
\new{It is worthy to mention that all the three tasks in this work described in Section~\ref{sec:problem_task} require the same parameters (user embeddings and item embeddings) for computing the relevance scores between users and items, and there are no task-specific parameters.}
Thus, we \new{omit $\theta^s$} and set the model parameter $\theta=\theta^c=\Theta=\llbracket\Theta^U, \Theta^I\rrbracket$. As a result, we generate all objectives using the unified parameter, which is the $\Theta$ including user and item embeddings. 

\new{For building the overall training objective, we use the weighted summation framework following~\cite{DBLP:journals/ior/Fishburn67, KIM_moop, Yang_Nature-Inspired} since this is the most common choice in optimizing multiple objectives.
The weighted summation is a convex combination of objectives and each single objective optimization determines one particular optimal solution point on the Pareto frontier. 
For ensuring the final solution is Pareto stationary (Pareto optimal under mild assumptions), we adopt Algorithm~\ref{alg: Frank-Wolfe Solver} to adaptively compute the scaling factors $\bm{\alpha}$ that satisfy the convex constraints in KKT conditions as aforementioned.
Thus, our overall training objective can be formulated as follows:}

\begin{equation}
\begin{aligned}
    \mathcal{L}= &\alpha^A\cdot \mathcal{L}^{Accuracy}+\sum_{i=1}^{m}\alpha^C_i\cdot \mathcal{L}^{C-Fair}_i+\sum_{j=1}^{n}\alpha^P_j\cdot \mathcal{L}^{P-Fair}_j,\\
    \text{s.t., }&  \alpha^A+\sum\limits_{i=1}^m\alpha^C_i+\sum\limits_{j=1}^n\alpha^P_j=1.\\
    \text{ }& \alpha^A\geq0, \alpha^C_i\geq0, \alpha^P_j\geq0, \forall i, j.
\label{eq: full_obj}
\end{aligned} 
\end{equation}
Here, $m$ and $n$ refer to the number of fairness constraints on the consumer and producer sides, respectively.

It is worth noticing that our proposed \textit{Multi-FR} framework does not rely on specific formulations of the loss functions or the model structures. Although the afore four disparity measures belong to the group fairness, one can also define any individual fairness objectives and scalably integrate them into our framework as long as they are differentiable.
\vspace{-0.1mm}

\begin{algorithm}[t!]
\caption{Multi-FR Framework }
\label{alg: Multi-FR}
\SetAlgoLined
Initialization( )\\
\For{$i\in 1, ..., t$}{
    Construct the individual objective $\mathcal{L}_i(\Theta)$ for each task\qquad \text{\small{\textit{(The objective for recommendation quality and the fairness constraints)}}}
}
\For{$epoch\in 1, ..., n_{epoch}$}{
    \For{$batch\in 1, ..., n_{batch}$}{
    Forward\_Passing( )\\
    \For{$i\in 1, ..., t$}{
        Compute the gradient for each objective: $\nabla_{\Theta}\mathcal{L}_i(\Theta)$\\
        Gradient\_Normalization( )\quad (\text{optional})
        
    }
    $\bm{\alpha}=(\alpha_1, ..., \alpha_t)\qquad\leftarrow\text{\new{Using Algorithm 1}}$\\
    Construct the single aggregated objective: $\mathcal{L}(\Theta)=\sum\limits_{i=1}^t\alpha_i\cdot\mathcal{L}_i(\Theta)$\\
    $\nabla_{\Theta}\mathcal{L}(\Theta)=\sum\limits_{i=1}^t\alpha_i\cdot\nabla_{\Theta}\mathcal{L}_i(\Theta)$\\
    Update $\Theta$
    }}
 \Return $\Theta$ that can lead to Pareto stationarity / Pareto optimality on quality and fairness
\end{algorithm}
\setlength{\textfloatsep}{8pt}

\subsection{Solution Selection}
\label{sub:solution_selection}
There is no consensus strategy on choosing one single Pareto optimal solution from a Pareto set since any solution in the Pareto set cannot strictly dominate the others.
To select a proper solution, we borrow the idea from one of the most well-known metrics in Theoretical Economics, the \textit{Least Misery Strategy}~\cite{DBLP:journals/mta/PessemierDM14}, for guiding us to select a ``fair'' solution for all the objectives. 
\new{~\citet{LinCPSXSZOJ19ParetoRecsys} also adopt this criteria to select the final Pareto optimal solution. 
The main idea is that we do not want the worst objective to be too bad.}

Motivated by the \textit{Least Misery Strategy}, our final solution aims to \new{choose the solution coming from the $q'^{th}$ round} that minimizes the highest loss value across all the objectives:
\setlength{\textfloatsep}{0.1pt}
\begin{align}
    \min\limits_{1\leq {q'}\leq q}\max\{\mathcal{L}^{q'}_1, \mathcal{L}^{q'}_2, ..., \mathcal{L}^{q'}_t\},
\label{eq:least_misery}
\end{align}
where $t$ is the total number of objectives in the model and $q$ is the total number of running rounds.
\new{The $t$ here can be defined as any positive integer depending on the number of objectives to model in regarding to the consumer-sided fairness and the producer-sided fairness}.
The superscript ${q'}$ indicates the ${q'}^{th}$ round. Therefore, given a generated Pareto frontier by running Algorithm~\ref{alg: Frank-Wolfe Solver} and Algorithm~\ref{alg: Multi-FR} for $q$ rounds ($q=5$) in our proposed model, we pick the solution coming from the round ${q'}$ with the minimum value of Eq.~\ref{eq:least_misery} as the final recommendation.

\section{Experiments}
In this section, we evaluate the proposed model and other baseline methods on three real-world datasets.

\subsection{Datasets}
The proposed model is evaluated on three real-world datasets from various domains with different sparsities: \textit{MovieLens100K}~\footnote{\href{https://grouplens.org/datasets/movielens/100k/}{https://grouplens.org/datasets/movielens/100k/}}, \textit{MovieLens1M}~\footnote{\href{https://grouplens.org/datasets/movielens/1m/}{https://grouplens.org/datasets/movielens/1m/}} \cite{DBLP:journals/tiis/HarperK16}, and \textit{FM-reduced}~\footnote{\href{http://ocelma.net/MusicRecommendationDataset/lastfm-360K.html}{http://ocelma.net/MusicRecommendationDataset/lastfm-360K.html}}. \textit{MovieLens100K} and \textit{MovieLens1M} are user-movie datasets collected from the \textit{MovieLens} website. These two datasets provide $100$ thousand and $1$ million user-movie interactions, respectively, with the user metadata (gender and age group) and movie genres. 
The \textit{FM-reduced} dataset is collected from the last.fm website, which contains the music listening records of $360$ thousand users along with the gender and age of users.
\new{The original version of this dataset is too large to run the majority of previously developed fairness-aware algorithms, as it would take a huge consumption both in time and space.
In order to make the experiments able to be conducted on all baselines, we reduce the size of the dataset by first randomly selecting 25,000 users.}
Under the implicit feedback setting, we keep those ratings no less than four (out of five) as positive feedback and treat all other ratings as missing entries for all datasets. 
To filter noisy data, we only keep the users with at least ten ratings and the items at least with five ratings. 

We adopt the age group strategy of the MovieLens dataset to split users into $7$ different age groups and the movies into $18$ different genres in all experiments. For all the datasets, we also group the items into $5$ different groups based on their popularity. For each user, we randomly split $70\%$, $10\%$, $20\%$ of the rated items as the training set, validation set, and testing set, respectively. The statistics for the datasets after preprocessing are shown in Table \ref{tab:data_statistics}. 

\begin{table}[t!]
 \centering
\caption{\label{tab:data_statistics}The statistics of datasets.}
\vspace{1mm}
 \begin{tabular}{lrrr}
 \toprule
   &   \textbf{MovieLens100K} & \textbf{MovieLens1M} & \textbf{\new{FM-reduced}}  \\
   \midrule
   
   \textbf{\#User} & 943 & 5,805 & 19,234\\
   \emph{\#female/\#male/\#age} & 273/670/7 & 1,655/4,150/7 & 4,022/15,212/7\\
   \midrule
   \textbf{\#Item} & 1,682 & 3,574 & 9,703 \\
   \emph{\#genre/\#popularity} & 18/5 & 18/5 & -/5\\
   \midrule
   \textbf{\#Interaction} & 100,458 & 678,740 & 1,049,322 \\
   \textbf{Density} & 6.33\% & 3.27\% & 0.56\%\\
\bottomrule
\end{tabular}
\end{table}
\setlength{\textfloatsep}{8pt}

\subsection{Evaluation Metrics}
In this section, we demonstrate our chosen metrics on the recommendation accuracy, fairness, and diversity. We adopt both self-defined metrics and commonly used measurements in academia. Our measurement of fairness and diversity covers the \textit{individual level}, \textit{group level}, and \textit{system level}.

\textit{Metrics for measuring  recommendation accuracy}:
\begin{itemize}[leftmargin=*]
\item \textbf{Recall@$K$}, which indicates the percentage of her rated items that appear in the top-$K$ recommended items. The greater the value of this metric, the higher the quality of the model.
\item \textbf{NDCG@$K$}, which is the normalized discounted cumulative gain at $K$, which takes the position of correctly recommended items into account. The greater the value of this metric, the higher the quality of the model.
\end{itemize}

\textit{Self-defined Metrics for measuring fairness}:
\begin{itemize}[leftmargin=*]
\item \textbf{Disparity$_u$} measures the unfairness on the user side, i.e. Eq.~\ref{eq:fairness_user}. The smaller the value of this metric, the higher the consumer-sided fairness of the model.
\item \textbf{Disparity$_i$} measures the unfairness on the item side, i.e. Eq.~\ref{eq:fairness_item}. The smaller the value of this metric, the higher the producer-sided fairness of the model.
\end{itemize}

\textit{General metrics for measuring fairness and diversity}:
\begin{itemize}[leftmargin=*]
\item \textbf{Gini Index} measures the inequality among values of a frequency distribution~\cite{GiniIndex}, e.g., numbers of occurrences (exposures) in the recommendation list. This measurement is at \textit{individual level}. Given a list of exposure of all items ($\mathcal{I}$) aggregated over all the recommendation lists, $l_e=[e_1, e_2, ..., e_{|\mathcal{I}|}]$, the Gini Index is calculated as below~\footnote{
\new{It is worthy to notice that, for Gini Index, there exist multiple alternative expressions~\cite{Gini-alter1, Gini-alter2}. 
Here we adopt the most widely used version as demonstrated in book~\cite{Gini} and paper~\cite{GeLGXLZP0GOZ21_longtermfairness}.  
The main difference between these definitions lies in whether to compute the coefficient with direct reference to the Lorenz curve~\cite{LorenzCurve}.}}:
\begin{equation}
\setlength\abovedisplayskip{2pt}
\setlength\belowdisplayskip{2pt}
    \text{Gini}(l_e)=\frac{1}{2|\mathcal{I}|^2\overline{e}}\sum_{i=1}^{|\mathcal{I}|}\sum_{j=1}^{|\mathcal{I}|}|e_i-e_j|,
\end{equation}
where $\overline{e}$ is the mean of all item exposures. The smaller the value of the Gini Index, the higher the fairness of the model.

\item \textbf{Popularity rate} computes the proportion of popular items in the recommendation list against the total number of items in the list, which can be regarded as a \textit{group-level} measurement. The smaller the value of the popularity rate, the higher the fairness of the model.
\item \textbf{Simpson's diversity index} \new{was introduced in 1949 by Edward H. Simpson to measure the degree of concentration when individuals are classified into types~\cite{Simpson'sDiversity}. 
This is suitable for the scenario of recommendation since items are generally categorized into different groups.
This metric was also adopted before by~\citet{Simpson_RecSys} for measuring the diversity in recommendation.}
Therefore, we employ this metric as a \textit{system-level} measurement of diversity that takes into account the number of groups present, as well as the relative abundance of each group.
Given a list of exposures of all items in the recommendation results and the group label of each, the Simpson's diversity index can be formulated as:
\begin{equation}
\setlength\abovedisplayskip{2pt}
\setlength\belowdisplayskip{2pt}
\text{Diversity}=1-(\frac{\sum_{i=1}^g n_i(n_i-1)}{N(N-1)})\,,
\end{equation}
where $g$ is the total number of groups, $n_i$ is the total number of items of group $i$, and $N$ is the total number of items of all groups. 
\new{This diversity index can also be interpreted as the probability that two randomly sampled items (without replacement) do not belong to the same group.}
The greater the value of this metric, the higher the diversity of the system.
\end{itemize}


\subsection{Method Studied}
\label{ref: method}
We choose three models as our recommendation backbones:
\begin{itemize}[leftmargin=*]
    \item \textbf{BPRMF}, Bayesian Personalized Ranking-based Matrix Factorization~\cite{rendle2012bpr}, which is a classic method for learning pairwise personalized rankings from user implicit feedback.
    \item \textbf{WRMF}, Weighted Regularized Matrix Factorization~\cite{4781121}, which minimizes the square error loss by assigning both observed and unobserved feedback with different confidential values based on matrix factorization.
    \item \textbf{NGCF}, Neural Graph Collaborative Filtering~\cite{DBLP:conf/sigir/Wang0WFC19}. This method integrates the user-item interactions into the embedding learning process, and exploits the graph structure by propagating embeddings on it to model the high-order connectivity.
\end{itemize}

We first adopt the above three backbones to learn the latent representation of users and items and obtain the relevance scores between them. Then we adopt the following three fairness-aware approaches on the top of the three backbones to achieve fair recommendation for a comparison with our proposed method.
\begin{itemize}[leftmargin=*]
    \item \textbf{FOEIR}, Fairness of Exposure in Rankings~\cite{DBLP:conf/kdd/SinghJ18}, which is a fairness-aware algorithm incorporating a standard linear program and the Birkhoff-von Neumann decomposition~\cite{birkhoff1967lattice}.
    \item \textbf{FairRec}, which is a two-sided fairness-aware method achieving envy-freeness up-to-one on the user side and exposure guarantee on the item side~\cite{PatroBGGC20FairRec}. It is motivated by the fair allocation~\cite{bouveret_chevaleyre_maudet_moulin_2016} and adopts the Greedy-Round-Robin algorithm~\cite{round-robin1,round-robin2} to allocate item candidates to users.
    \item \new{\textbf{TFROM}, which is a two-sided fairness-aware method ensuring individual fairness on both consumers and producers~\cite{Wu0XT21TFROM}. It also uses scheduling algorithm for conducting the recommendation. }
\end{itemize}

\new{Here, we summarize how the fairness criteria are defined in these fairness-aware recommendation approaches.
\textbf{FOEIR} is a ranking model for search, thus it does not model consumer-sided fairness.
It considers three types of fairness on the producer side: (1) Demographic Parity: $E(G_1)=E(G_2)$, which ensures that the exposure (E) obtained by different demographic groups be equal. (2) Disparate Treatment: $\frac{E(G_1)}{U(G_1)}=\frac{E(G_2)}{U(G_2)}$, which ensures that the exposure ($E$) should be proportional to the utility ($U$) of members in each group. (3) Disparate Impact: $\frac{CTR(G_1)}{U(G_1)}=\frac{CTR(G_2)}{U(G_2)}$, which cares more about the final effect and ensures that the click-through rate ($CTR$) of items belonging to different groups should be proportional to the utility ($U$).
In our experiment, we choose the Demographic Parity as the fairness constraint of FOEIR.
\textbf{FairRec} uses the \textit{envy-freeness-up-to-one (EF1)} good to define the individual fairness on the consumer side: $v_{u_1}(\mathcal{A}_{u_1})\geq v_{u_1}(\mathcal{A}_{u_2}\backslash{p})$.
This definition comes from the fair allocation, which is a sub-field of economics.
Here $v_{u_1}(\mathcal{A}_{u_1})$ denotes the amount how $u_1$ values his obtained allocation (recommended items) $\mathcal{A}_{u_1}$, and $p$ is any item in another user $u_2$'s allocation $\mathcal{A}_{u_2}$.
For the producer-sided fairness, the authors define it as ensuring a (self-defined) minimum threshold of exposure for all items.
\textbf{TFROM} still defines an individual fairness on the consumer side as ensuring the NDCG values of any two users are equal: $NDCG_{u_1} = NDCG_{u_2}$. 
The authors define two producer-sided fairness criteria both at an individual level: (1) Uniform Fairness: $\frac{E(p_1)}{|I_{p_1}|}=\frac{E(p_2)}{|I_{p_2}|}$, which ensures that the exposure of each individual producer should be proportional to the number of items ($|I|$) she offers. (2) Quality Weighted Fairness: $\frac{E(p_1)}{Q(I_{p_1})}=\frac{E(p_2)}{Q(I_{p_2})}$, which ensures that the exposure of each individual producer should be proportional to the quality ($Q$) of items she offers.
We summarize all these fairness criteria in Table.~\ref{table:prior_def}.
}

\begin{table*}[t!]
\centering
\caption{\label{table:prior_def}\new{Definitions of fairness in the three fairness-aware approaches we selected for comparison. The definitions of notations are elaborated in Section.~\ref{ref: method}.}}
    \centering
    \begin{tabular}{c|c|c}
    \toprule
      Approach  & Consumer-sided Fairness & Producer-sided Fairness \\
      \midrule
\multirow{4}{*}{FOEIR~\cite{DBLP:conf/kdd/SinghJ18}} & \multirow{4}{*}{N.A.} & \textbf{Group Fairness}\\
& & Demographic Parity: $E(G_1)=E(G_2)$\\
& & Disparate Treatment: $\frac{E(G_1)}{U(G_1)}=\frac{E(G_2)}{U(G_2)}$ \\
& & Disparate Impact: $\frac{CTR(G_1)}{U(G_1)}=\frac{CTR(G_2)}{U(G_2)}$\\
\midrule
\multirow{2}{*}{FairRec~\cite{PatroBGGC20FairRec}} & \textbf{Individual Fairness} & \textbf{Group Fairness}\\
& EF1: $v_{u_1}(\mathcal{A}_{u_1})\geq v_{u_1}(\mathcal{A}_{u_2}\backslash{p})$ & Guarantee a threshold of exposure for all producers\\
\midrule
\multirow{3}{*}{TFROM~\cite{Wu0XT21TFROM}} & \textbf{Individual Fairness} & \textbf{Individual Fairness} \\
& \multirow{2}{*}{$NDCG_{u_1} = NDCG_{u_2}$} & Uniform Fairness: $\frac{E(p_1)}{|I_{p_1}|}=\frac{E(p_2)}{|I_{p_2}|}$\\
& & Quality Weighted Fairness: $\frac{E(p_1)}{Q(I_{p_1})}=\frac{E(p_2)}{Q(I_{p_2})}$\\
\bottomrule
    \end{tabular}
\end{table*}

Lastly, we adopt our proposed \textbf{MultiFR} method on the top of the BPRMF, WRMF, and NGCF to form our final model. Our method allows the weights on different objectives to be adaptively learned during the training process with the model embeddings.
\begin{table*}[t!]
\caption{\label{tab:overall_performance_ml100k}Summary of the performance on \textbf{MovieLens100K}. We evaluate for recommendation accuracy (\textit{Recall} and \textit{NDCG}) and fairness (\textit{Disparity$_u$}, \textit{Disparity$_i$}, \textit{Gini}, \textit{Popularity rate}, and \textit{Diversity}), where $K$ is the length of the recommendation list. A metric followed by ``$\uparrow$'' means ``the larger, the better'', while a metric followed by ``$\downarrow$'' means ``the smaller, the better''. The fairness constraints are specified using the \textit{``gender''} on the consumer side and the \textit{``genre''} on the producer side. In each block, the paired t-test between the second best method and the best method on each metric is significant at $p\leq0.01$.}
\vspace{1mm}
\centering
\resizebox{\textwidth}{!}{
\begin{tabular}{l|p{1.5cm}<{\centering}p{1.5cm}<{\centering}|p{1.5cm}<{\centering}p{1.5cm}<{\centering}p{1.5cm}<{\centering}p{1.5cm}<{\centering}p{1.5cm}<{\centering}}
\toprule
Model & \multicolumn{1}{c}{Recall@K $\uparrow$} & \multicolumn{1}{c|}{NDCG@K $\uparrow$} & \multicolumn{1}{c}{Disparity$_u$ $\downarrow$} & \multicolumn{1}{c}{Disparity$_i$ $\downarrow$} & \multicolumn{1}{c}{Gini $\downarrow$} & \multicolumn{1}{c}{\makecell{Popularity\\ rate} $\downarrow$} &  \multicolumn{1}{c}{Diversity $\uparrow$}\\
\midrule
\multicolumn{8}{c}{K=10}\\
\midrule
BPRMF & \textbf{0.2152} & \textbf{0.2637} & 1.3580 & 1.2131 & 0.6919 & 0.8996 & 0.1838\\
BPRMF-FOEIR & 0.1968 & 0.2473 & 1.2103 & 1.0023 & 0.6558 & 0.8351 & 0.2817\\
BPRMF-FairRec & 0.2049 & 0.2495 & 1.3627 & 0.9357 & 0.6325 & 0.8214 & 0.2916\\
BPRMF-TFROM & 0.2037 & 0.2481 & \underline{1.1703} & \underline{0.9223} & \underline{0.6138} & \underline{0.8172} & \underline{0.2968}\\
\textbf{BPRMF-MultiFR} & \underline{0.2055} & \underline{0.2516} & \textbf{0.9877} & \textbf{0.8235} & \textbf{0.6027} & \textbf{0.8032} & \textbf{0.3029}\\
\midrule
WRMF & \textbf{0.2166}  & \textbf{0.2748} & 1.3753 & 1.2215 & 0.7278 & 0.9256 & 0.1391\\
WRMF-FOEIR & 0.2107 & 0.2694 & 1.2533 & 1.1185 & 0.7152 & 0.8574 & 0.2478\\
WRMF-FairRec & 0.2140 & \underline{0.2735} & 1.3624 & 0.9626 & 0.6954 & 0.8315 & 0.2621\\
WRMF-TFROM & 0.2128 & 0.2707 & \underline{1.2213} & \underline{0.9266} & \underline{0.6424} & \underline{0.8106} & \underline{0.2754}\\
\textbf{WRMF-MultiFR} & \underline{0.2156} & 0.2733 & \textbf{0.9997} & \textbf{0.8672} & \textbf{0.6239} & \textbf{0.8021} & \textbf{0.3045}\\
\midrule
NGCF & \textbf{0.2275} & \textbf{0.2855} & 1.4226 & 1.2333 & 0.7526 & 0.9621 & 0.1023\\
NGCF-FOEIR & 0.2242 & 0.2705 & 1.2624 & 1.1388 & 0.7239 & 0.8627 & 0.2282\\
NGCF-FairRec & \underline{0.2252} & \underline{0.2776} & 1.3526 & 1.0221 & 0.7027 & 0.8410 & 0.2518\\
NGCF-TFROM & 0.2219 & 0.2756 & \underline{1.2317} & \underline{1.0005} & \underline{0.6811} & \underline{0.8313} & \underline{0.2724}\\
\textbf{NGCF-MultiFR} & 0.2245 & 0.2752 & \textbf{1.0232} & \textbf{0.9862} & \textbf{0.6428} & \textbf{0.8213} & \textbf{0.3042}\\
\midrule
\multicolumn{8}{c}{K=20}\\
\midrule
BPRMF & \textbf{0.3273} & \textbf{0.2848} & 2.8215 & 1.2237 & 0.6840 & 0.8630 & 0.2367\\
BPRMF-FOEIR & 0.3107 & 0.2734 & 2.2250 & 1.0011 & 0.6540 & 0.7934 & 0.3381\\
BPRMF-FairRec & 0.3204 & 0.2801 & 2.5129 & 1.0001 & 0.6459 & \underline{0.7589} & \underline{0.3426}\\
BPRMF-TFROM & 0.3134 & 0.2782 & \underline{2.0781} & \underline{0.9283} & \underline{0.6271} & 0.7602 & 0.3397\\
\textbf{BPRMF-MultiFR} & \underline{0.3210} & \underline{0.2833} & \textbf{1.9234} & \textbf{0.8826} & \textbf{0.6011} & \textbf{0.7552} & \textbf{0.3625}\\
\midrule
WRMF & \textbf{0.3305} & \textbf{0.2964} & 3.0307 & 1.2382 & 0.7223 & 0.8953 & 0.1905\\
WRMF-FOEIR & 0.3221 & 0.2896 & 2.5632 & 1.1296 & 0.6916 & 0.8051 & 0.3217\\
WRMF-FairRec & 0.3227 & \underline{0.2925} & 2.8691 & 1.0214 & 0.6729 & 0.7926 & 0.3281\\
WRMF-TFROM & 0.3209 & 0.2912 & \underline{2.4253} & \underline{0.9721} & \underline{0.6462} & \underline{0.7891} & \underline{0.3402}\\
\textbf{WRMF-MultiFR} & \underline{0.3255} & 0.2921 & \textbf{2.0913} & \textbf{0.9023} & \textbf{0.6124} & \textbf{0.7889} & \textbf{0.3588}\\
\midrule
NGCF & \textbf{0.3431} & \textbf{0.3022} & 3.2431 & 1.2347 & 0.7728 & 0.9233 & 0.1525\\
NGCF-FOEIR & 0.3259 & 0.2953 & 2.6877 & 1.1465 & 0.6966 & 0.8234 & 0.3029\\
NGCF-FairRec & \underline{0.3327} & 0.2996 & 2.9162 & 1.1056 & 0.6735 & 0.8134 & 0.3194\\
NGCF-TFROM & 0.3317 & 0.2989 & \underline{2.4726} & \underline{1.0314} & \underline{0.6534} & \underline{0.8108} & \underline{0.3356}\\
\textbf{NGCF-MultiFR} & 0.3286 & \underline{0.3000} & \textbf{2.2421} & \textbf{0.9928} & \textbf{0.6421} & \textbf{0.8001} & \textbf{0.3429}\\
\bottomrule
\end{tabular}
}
\end{table*}

\begin{table*}[t!]
\caption{\label{tab:overall_performance_ml1m}Summary of the performance on \textbf{MovieLens1M}. We evaluate for recommendation accuracy (\textit{Recall} and \textit{NDCG}) and fairness (\textit{Disparity$_u$}, \textit{Disparity$_i$}, \textit{Gini}, \textit{Popularity rate}, and \textit{Diversity}), where $K$ is the length of the recommendation list. A metric followed by ``$\uparrow$'' means ``the larger, the better'', while a metric followed by ``$\downarrow$'' means ``the smaller, the better''. The fairness constraints are specified using the \textit{``gender''} on the consumer side and the \textit{``genre''} on the producer side. In each block, the paired t-test between the second best method and the best method on each metric is significant at $p\leq0.01$.}
\vspace{1mm}
\centering
\resizebox{\textwidth}{!}{
\begin{tabular}{l|p{1.5cm}<{\centering}p{1.5cm}<{\centering}|p{1.5cm}<{\centering}p{1.5cm}<{\centering}p{1.5cm}<{\centering}p{1.5cm}<{\centering}p{1.5cm}<{\centering}}
\toprule
Model & \multicolumn{1}{c}{Recall@K $\uparrow$} & \multicolumn{1}{c|}{NDCG@K $\uparrow$} & \multicolumn{1}{c}{Disparity$_u$ $\downarrow$} & \multicolumn{1}{c}{Disparity$_i$ $\downarrow$} & \multicolumn{1}{c}{Gini $\downarrow$} & \multicolumn{1}{c}{\makecell{Popularity\\ rate} $\downarrow$} &  \multicolumn{1}{c}{Diversity $\uparrow$}\\
\midrule
\multicolumn{8}{c}{K=10}\\
\midrule
BPRMF & \textbf{0.1462} & \textbf{0.2360} & 1.5225 & 1.2648 & 0.7586 & 0.9326 & 0.1264\\
BPRMF-FOEIR & 0.1425 & 0.2318 & 1.4263 & 1.2624 & 0.7171 & 0.8530 & 0.2545\\
BPRMF-FairRec & 0.1453 & \underline{0.2344} & 1.4927 & 1.0826 & 0.6927 & 0.8531 & 0.2637\\
BPRMF-TFROM & 0.1437 & 0.2332 & \underline{1.3782} & \underline{0.9263} & \underline{0.6918} & \underline{0.8346} & \underline{0.2812}\\
\textbf{BPRMF-MultiFR} & \underline{0.1458} & 0.2333 & \textbf{1.0235} & \textbf{0.8716} & \textbf{0.6825} & \textbf{0.8214} & \textbf{0.3023}\\
\midrule
WRMF & \textbf{0.1681} & \textbf{0.2850} & 2.1773 & 1.3125 & 0.7720 & 0.9921 & 0.0157\\
WRMF-FOEIR & 0.1646 & 0.2809 & 2.1750 & 1.3128 & 0.7710 & 0.9244 & 0.0873\\
WRMF-FairRec & \underline{0.1661} & \underline{0.2846} & 2.1023 & 1.1352 & 0.7241 & 0.9027 & 0.1015\\
WRMF-TFROM & 0.1654 & 0.2829 & \underline{1.9273} & \underline{1.1044} & \underline{0.7172} & \underline{0.8823} & \underline{0.1025}\\
\textbf{WRMF-MultiFR} & 0.1644 & 0.2811 & \textbf{1.6523} & \textbf{0.9726} & \textbf{0.7032} & \textbf{0.8527} & \textbf{0.1029}\\
\midrule
NGCF & \textbf{0.1782} & \textbf{0.2852} & 2.5632 & 1.3527 & 0.8010 & 0.9935 & 0.0032\\
NGCF-FOEIR & 0.1762 & 0.2834 & 2.3345 & 1.3189 & 0.7786 & 0.9305 & 0.0127\\
NGCF-FairRec & \underline{0.1774} & \underline{0.2848} & 2.5413 & 1.2635 & 0.7309 & 0.9135 & 0.1000\\
NGCF-TFROM & 0.1767 & 0.2833 & \underline{2.1311} & \underline{1.1472} & \underline{0.7222} & \underline{0.8873} & \underline{0.1124}\\
\textbf{NGCF-MultiFR} & 0.1724 & 0.2829 & \textbf{1.8528} & \textbf{1.1125} & \textbf{0.7152} & \textbf{0.8734} & \textbf{0.1320}\\
\midrule
\multicolumn{8}{c}{K=20}\\
\midrule
BPRMF & \textbf{0.2287} & \textbf{0.2438} & 3.2123 & 1.2638 & 0.7512 & 0.9047 & 0.1743\\
BPRMF-FOEIR & 0.2220 & 0.2384 & 2.8707 & 1.2660 & 0.7034 & \underline{0.8075} & 0.3183\\
BPRMF-FairRec & \underline{0.2280} & \underline{0.2425} & 2.9012 & 1.1109 & 0.6931 & 0.8123 & 0.3237\\
BPRMF-TFROM & 0.2272 & 0.2419 & \underline{2.7100} & \underline{1.0019} & \underline{0.6846} & 0.8087 & \underline{0.3308}\\
\textbf{BPRMF-MultiFR} & 0.2252 & 0.2424 & \textbf{2.5972} & \textbf{0.8241} & \textbf{0.6728} & \textbf{0.8027} & \textbf{0.3426}\\
\midrule
WRMF & \textbf{0.2525} & \textbf{0.2859} & 3.9079 & 1.3105 & 0.7579 & 0.9808 & 0.0377\\
WRMF-FOEIR & 0.2469 & 0.2804 & 3.8470 & 1.3105 & 0.7534 & 0.8967 & 0.1854\\
WRMF-FairRec & \underline{0.2502} & \underline{0.2851} & 3.8721 & 1.2358 & 0.7129 & 0.8749 & 0.2203\\
WRMF-TFROM & 0.2482 & 0.2839 & \underline{3.7247} & \underline{1.1392} & \underline{0.7072} & \underline{0.8562} & \underline{0.2210}\\
\textbf{WRMF-MultiFR} & 0.2470 & 0.2832 & \textbf{2.9341} & \textbf{1.0826} & \textbf{0.6923} & \textbf{0.8231} & \textbf{0.2239}\\
\midrule
NGCF & \textbf{0.2633} & \textbf{0.2936} & 4.1124 & 1.3469 & 0.7992 & 0.9922 & 0.0123\\
NGCF-FOEIR & 0.2574 & \underline{0.2890} & 3.8728 & 1.3098 & 0.7842 & 0.9231 & 0.1026\\
NGCF-FairRec & \underline{0.2591} & 0.2856 & 3.8927 & 1.2533 & 0.7542 & 0.8862 & 0.1224\\
NGCF-TFROM & 0.2580 & 0.2819 & \underline{3.5820} & \underline{1.1301} & \underline{0.7278} & \underline{0.8635} & \underline{0.1374}\\
\textbf{NGCF-MultiFR} & 0.2588 & 0.2844 & \textbf{3.0375} & \textbf{1.1057} & \textbf{0.6955} & \textbf{0.8562} & \textbf{0.1524}\\
\bottomrule
\end{tabular}
}
\end{table*}

\begin{table*}[t!]
\caption{\label{tab:overall_performance_lastfm}Summary of the performance on \textbf{FM-reduced}. We evaluate for recommendation accuracy (\textit{Recall} and \textit{NDCG}) and fairness (\textit{Disparity$_u$}, \textit{Disparity$_i$}, \textit{Gini}, \textit{Popularity rate}, and \textit{Diversity}), where $K$ is the length of the recommendation list. A metric followed by ``$\uparrow$'' means ``the larger, the better'', while a metric followed by ``$\downarrow$'' means ``the smaller, the better''. The fairness constraints are specified using the \textit{``age''} on the consumer side and the \textit{``popularity''} on the producer side. 
In each block, the paired t-test between the second best method and the best method on each metric is significant at $p\leq0.01$.}
\vspace{1mm}
\centering
\resizebox{\textwidth}{!}{
\begin{tabular}{l|p{1.5cm}<{\centering}p{1.5cm}<{\centering}|p{1.5cm}<{\centering}p{1.5cm}<{\centering}p{1.5cm}<{\centering}p{1.5cm}<{\centering}p{1.5cm}<{\centering}}
\toprule
Model & \multicolumn{1}{c}{Recall@K $\uparrow$} & \multicolumn{1}{c|}{NDCG@K $\uparrow$} & \multicolumn{1}{c}{Disparity$_u$ $\downarrow$} & \multicolumn{1}{c}{Disparity$_i$ $\downarrow$} & \multicolumn{1}{c}{Gini $\downarrow$} & \multicolumn{1}{c}{\makecell{Popularity\\ rate} $\downarrow$} &  \multicolumn{1}{c}{Diversity $\uparrow$}\\
\midrule
\multicolumn{8}{c}{K=10}\\
\midrule
BPRMF & \textbf{0.1248} & \textbf{0.1671} & 1.3277 & 1.3099 & 0.8136 & 0.9893 & 0.0211\\
BPRMF-FOEIR & \underline{0.1245} & \underline{0.1669} & 1.2746 & 1.2801 & 0.8064 & 0.9733 & 0.0732\\
BPRMF-FairRec & 0.1240 & 0.1659 & 1.3320 & 1.1511 & 0.7826 & 0.9625 & 0.1027\\
BPRMF-TFROM & 0.1227 & 0.1632 & \underline{1.1371} & \underline{1.1023} & \underline{0.7644} & \underline{0.9285} & \underline{0.1266}\\
\textbf{BPRMF-MultiFR} & 0.1209 & 0.1592 & \textbf{1.0288} & \textbf{0.9323} & \textbf{0.7514} & \textbf{0.9023} & \textbf{0.1674}\\
\midrule
WRMF & \textbf{0.1326} & \textbf{0.1828} & 1.6231 & 1.5135 & 0.8523 & 0.9905 & 0.0104\\
WRMF-FOEIR & \underline{0.1323} & \underline{0.1825} & 1.5268 & 1.3687 & 0.8271 & 0.9625 & 0.1162\\
WRMF-FairRec & 0.1322 & 0.1820 & 1.5826 & 1.2231 & 0.8038 & 0.9699 & 0.1008\\
WRMF-TFROM & 0.1320 & 0.1817 & \underline{1.3917} & \underline{1.1343} & \underline{0.7996} & \underline{0.9555} & \underline{0.1229}\\
\textbf{WRMF-MultiFR} & 0.1301 & 0.1784 & \textbf{1.1273} & \textbf{1.0824} & \textbf{0.7823} & \textbf{0.9275} & \textbf{0.1462}\\
\midrule
NGCF & \textbf{0.1452} & \textbf{0.1923} & 1.8231 & 1.8326 & 0.9006 & 0.9932 & 0.0096\\
NGCF-FOEIR & 0.1428 & 0.1899 & 1.6092 & 1.5247 & 0.8725 & 0.9755 & 0.0976\\
NGCF-FairRec & \underline{0.1435} & \underline{0.1901} & 1.6235 & 1.3825 & 0.8522 & 0.9826 & 0.0927\\
NGCF-TFROM & 0.1430 & 0.1892 & \underline{1.4326} & \underline{1.3728} & \underline{0.8364} & \underline{0.9678} & \underline{0.1072}\\
\textbf{NGCF-MultiFR} & 0.1426 & 0.1888 & \textbf{1.2526} & \textbf{1.1081} & \textbf{0.8002} & \textbf{0.9388} & \textbf{0.1388}\\
\midrule
\multicolumn{8}{c}{K=20}\\
\midrule
BPRMF & \textbf{0.1904} & \textbf{0.1892} & 1.3658 & 1.3103 & 0.8161 & 0.9792 & 0.0407\\
BPRMF-FOEIR & 0.1899 & \underline{0.1888} & 1.2541 & 1.3746 & 0.8006 & 0.9033 & 0.1752\\
BPRMF-FairRec & \underline{0.1902} & 0.1872 & 1.3435 & 1.1627 & 0.7519 & 0.8892 & 0.2016\\
BPRMF-TFROM & 0.1871 & 0.1836 & \underline{1.2762} & \underline{1.1517} & \underline{0.7498} & \underline{0.8554} & \underline{0.2263}\\
\textbf{BPRMF-MultiFR} & 0.1853 & 0.1726 & \textbf{0.9999} & \textbf{0.9083} & \textbf{0.7426} & \textbf{0.8388} & \textbf{0.2515}\\
\midrule
WRMF & \textbf{0.2104} & \textbf{0.2031} & 1.6127 & 1.6852 & 0.8627 & 0.9889 & 0.0214\\
WRMF-FOEIR & 0.2096 & \underline{0.2008} & 1.5179 & 1.4920 & 0.8489 & 0.9258 & 0.1237\\
WRMF-FairRec & \underline{0.2100} & 0.1984 & 1.5726 & 1.2338 & 0.8023 & 0.9022 & 0.1539\\
WRMF-TFROM & 0.2087 & 0.1976 & \underline{1.3744} & \underline{1.2076} & \underline{0.7926} & \underline{0.8825} & \underline{0.1627}\\
\textbf{WRMF-MultiFR} & 0.2062 & 0.1954 & \textbf{1.0862} & \textbf{1.0927} & \textbf{0.7782} & \textbf{0.8526} & \textbf{0.2073}\\
\midrule
NGCF & \textbf{0.2247} & \textbf{0.2258} & 1.7923 & 1.9349 & 0.9138 & 0.9905 & 0.0102\\
NGCF-FOEIR & 0.2229 & \underline{0.2206} & 1.6138 & 1.5562 & 0.8623 & 0.9429 & 0.1058\\
NGCF-FairRec & \underline{0.2236} & 0.2197 & 1.6282 & 1.3791 & 0.8425 & 0.9273 & 0.1286\\
NGCF-TFROM & 0.2225 & 0.2188 & \underline{1.4131} & \underline{1.2231} & \underline{0.8123} & \underline{0.9076} & \underline{0.1486}\\
\textbf{NGCF-MultiFR} & 0.2197 & 0.2164 & \textbf{1.2830} & \textbf{1.1001} & \textbf{0.7849} & \textbf{0.8862} & \textbf{0.1848}\\
\bottomrule
\end{tabular}
}
\end{table*}



\subsection{Experiment Settings}
In the experiments, we optimize all models using the Adam optimizer with the Xavier initialization~\cite{DBLP:journals/jmlr/GlorotB10}. The embedding size is fixed to $50$ and the batch size to $1024$ for all baseline models. The learning rate and the regularization hyper-parameter are selected from $ \{1e^{-1}, 1e^{-2}, 1e^{-3}, 1e^{-4}, 1e^{-5} \} $. The patience parameter $\gamma$ is selected from $\{0.5, 0.6, 0.7, 0.8\}$. The smooth temperature in SmoothRank is selected from $\{1e^{-1}, 1e^{-2}, 1e^{-3}\}$. The $K$ value in NDCG@$K$ used for computing the consumer-side fairness described in Section~\ref{subsub:consumer-sided fairness} is set as $20$. For all the datasets, we randomly sample one unobserved item as the negative sample for each user to speed up the training process. Further, for the FOEIR model, since it requires to solve a linear program with size $|\mathcal{I}|\times|\mathcal{I}|$ for each consumer with huge computational costs, we rerank the top-$100$ items from the base model then select the new top-$K$ ($K<100$) as the final recommendation. 
Early stopping strategy is performed, i.e., permutate stopping if Recall@$20$ on the validation data does not increase for 50 successive evaluation steps, for which the evaluation process is conducted for every five epochs. All experiments are conducted with PyTorch running on GPU machines (Nvidia Tesla P100).

\begin{figure}[p]
\centering
\subfigure[\textbf{MovieLens100K}]{
\begin{minipage}[t]{1.0\linewidth}
\centering
\includegraphics[width=\linewidth]{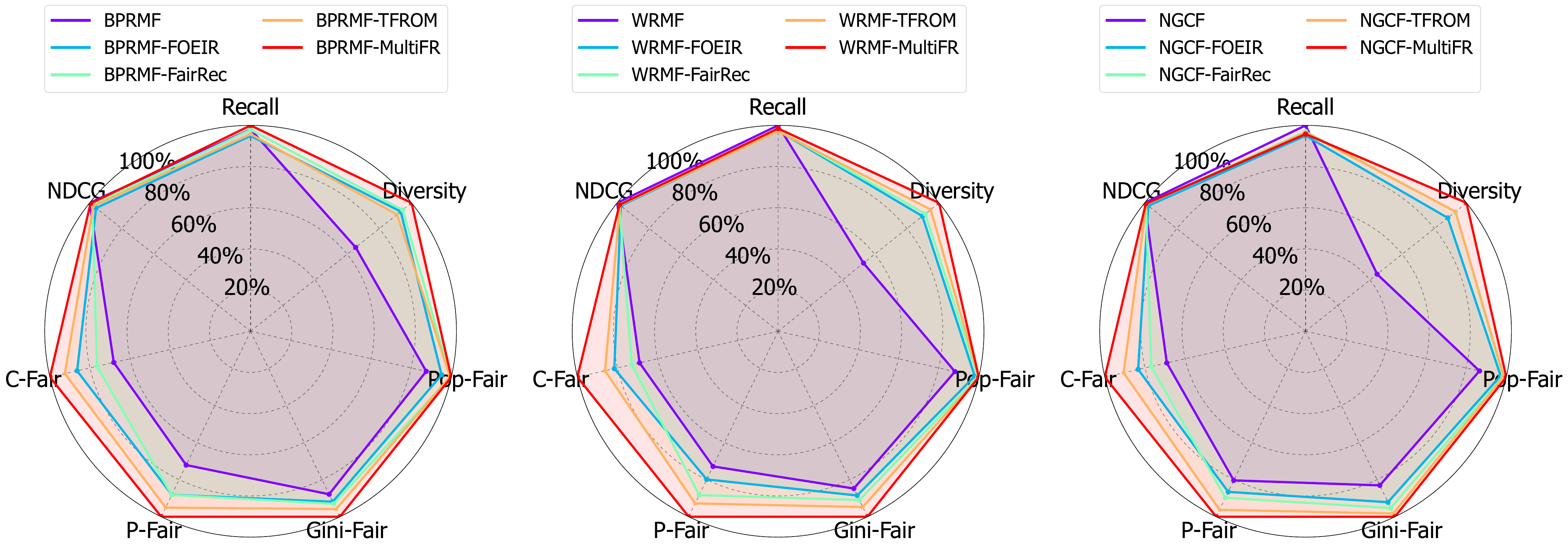}
\end{minipage}
}
\subfigure[\textbf{MovieLens1M}]{
\begin{minipage}[t]{1.0\linewidth}
\centering
\includegraphics[width=\linewidth]{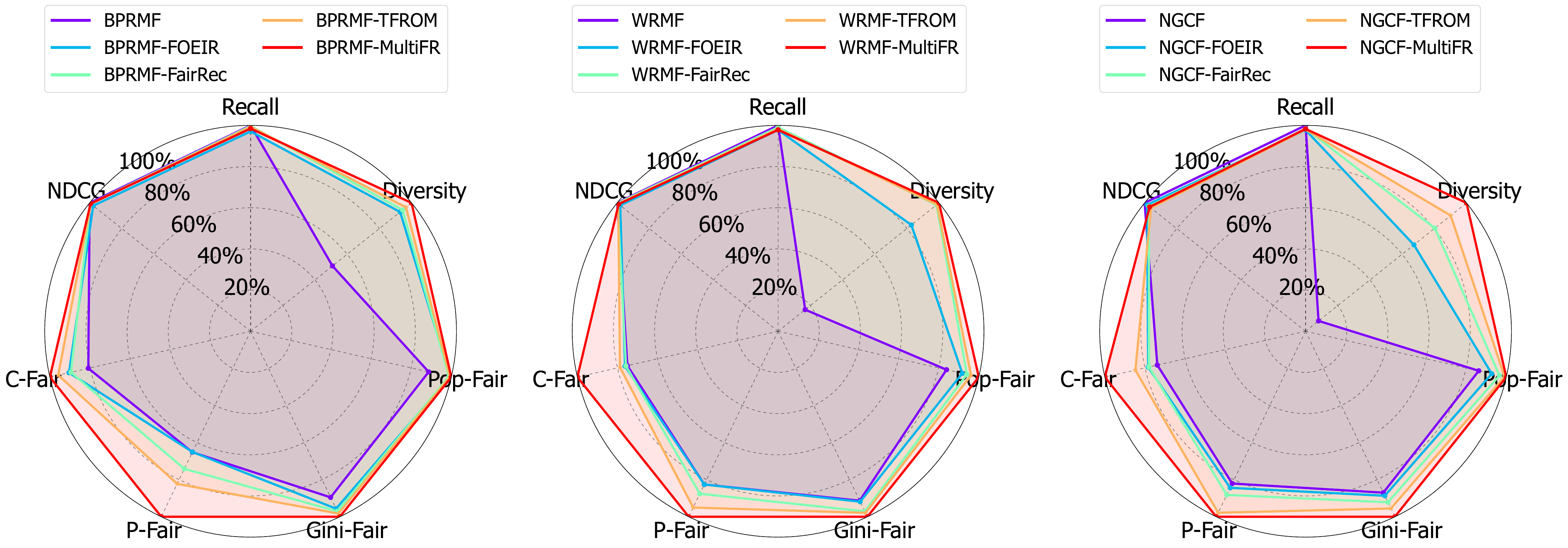}
\end{minipage}
}
\subfigure[\textbf{FM-reduced}]{
\begin{minipage}[t]{1.0\linewidth}
\centering
\includegraphics[width=\linewidth]{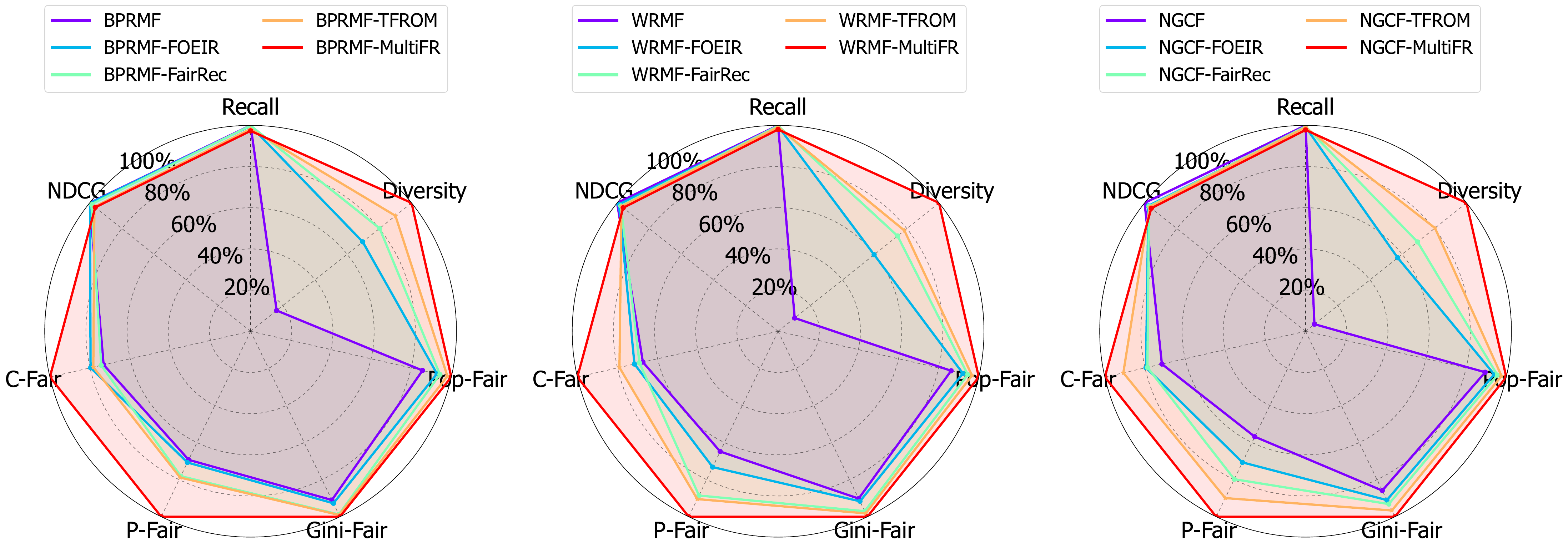}
\end{minipage}
}
\centering
\caption{Relative performance achievement compared to the best on each metric given the same backbone and dataset. All the metrics are computed at position 20 (i.e., \textit{Metric@20}). For \textit{Recall}, \textit{NDCG}, and \textit{Diversity} (the larger, the better), we divide the value achieved by each method by the best one to indicate the relative performance on the recommendation accuracy of each method compared to the best one. For other metrics (the smaller, the better), we first compute the reciprocal of each value to indicate the fairness from different perspectives, followed by adopting the same operation we conduct before. Thus, the greater the percentage value in these plots, the better the approach on that measurement.}
\label{pic:relevance_performance}
\end{figure}

\subsection{Experimental Results and Analysis}
\subsubsection{Overall Performance Comparison}
The overall experiments on three datasets are reported in Table~\ref{tab:overall_performance_ml100k}, Table~\ref{tab:overall_performance_ml1m}, and Table~\ref{tab:overall_performance_lastfm}, respectively.
\new{In each block, bold scores are the best for each metric, while underlined scores are the second best.}

Our model achieves obvious and significant improvements regarding all the fairness and diversity metrics. For instance, on the \textit{MovieLens100K} dataset, considering the top-10 recommendation, BPRMF-MultiFR reduces the disparity on the user side by 27.27\% and 32.12\% on the item side compared with the BPRMF base model. WRMF-MultiFR reduces the Gini index and Popularity rate by 13.04\% and 13.34\%, respectively. And NGCF-MultiFR model improves the system's diversity from 0.1023 to 0.3042, which is a rather great enhancement. The biggest improvement of the diversity metric is on the \textit{FM-reduced} dataset, where the diversity measure is improved from 0.0211 to 0.1674 by BPRMF-MultiFR compared with the corresponding base model. WRMF-MultiFR and NGCF-MultiFR also largely enhance the diversity by a large margin. Furthermore, compared with other state-of-the-art fair ranking methods, Multi-FR can still consistently achieve better fairness measures on both sides. 

We also observe a conflict between the recommendation accuracy and fairness. For instance, NGCF achieves the highest accuracy regarding Recall and NDCG on three datasets; however, its recommendation is the least fair and diverse compared to other models. FOEIR, FairRec, and TFROM achieve better fairness by re-ranking the recommendation list based on the relevance scores obtained from the base models; however, the original ranking order is disrupted, leading to the accuracy drop. 
\new{Different from prior post-processing methods that require to obtain the relevance scores between users and items beforehand, our Multi-FR is an in-processing method that ensures both the recommendation accuracy and multi-stakeholder fairness in an end-to-end way.} 
Based on the experimental results, Multi-FR can balance the recommendation accuracy and fairness well by largely improving the fairness and diversity with little drop in the accuracy. 
For instance, concerning Recall@20, NGCF-MultiFR only has a drop of 4.23\%, 1.71\%, and 2.23\% on three datasets, respectively, compared to the original NGCF model. 
Considering the large magnitude of fairness and diversity improvements, we denote this accuracy drop is relatively small.

In order to show the capability of our proposed method for balancing the recommendation accuracy and fairness more clearly, we display the radar plots in Fig.~\ref{pic:relevance_performance}, where each sub-plot compares the FOEIR, FairRec, TFROM, and Multi-FR on a specific backbone. For Recall, NDCG, and Diversity, we divide the value achieved by each method by the highest value to indicate how much percentage different methods can reach compared to the best one given the same backbone and dataset. For other metrics which are the smaller the better, we first compute a reciprocal of those values, indicating the fairness on different perspectives, i.e., Consumer (C), Producer (P), Gini, Popularity (Pop). Then we adopt a similar way to divide the value achieved by each method by the highest value obtained across all approaches to determine the relative scale that each method may attain in comparison to the best. As illustrated in Fig.~\ref{pic:relevance_performance}, the Multi-FR method outperforms all other approaches on all fairness metrics and produces nearly identical recommendation accuracy, regardless of the backbone and dataset chosen.


\begin{figure}[t]
\centering
\subfigure[\textbf{MovieLens100K}]{
\begin{minipage}[t]{1.0\linewidth}
\centering
\includegraphics[width=4.0in]{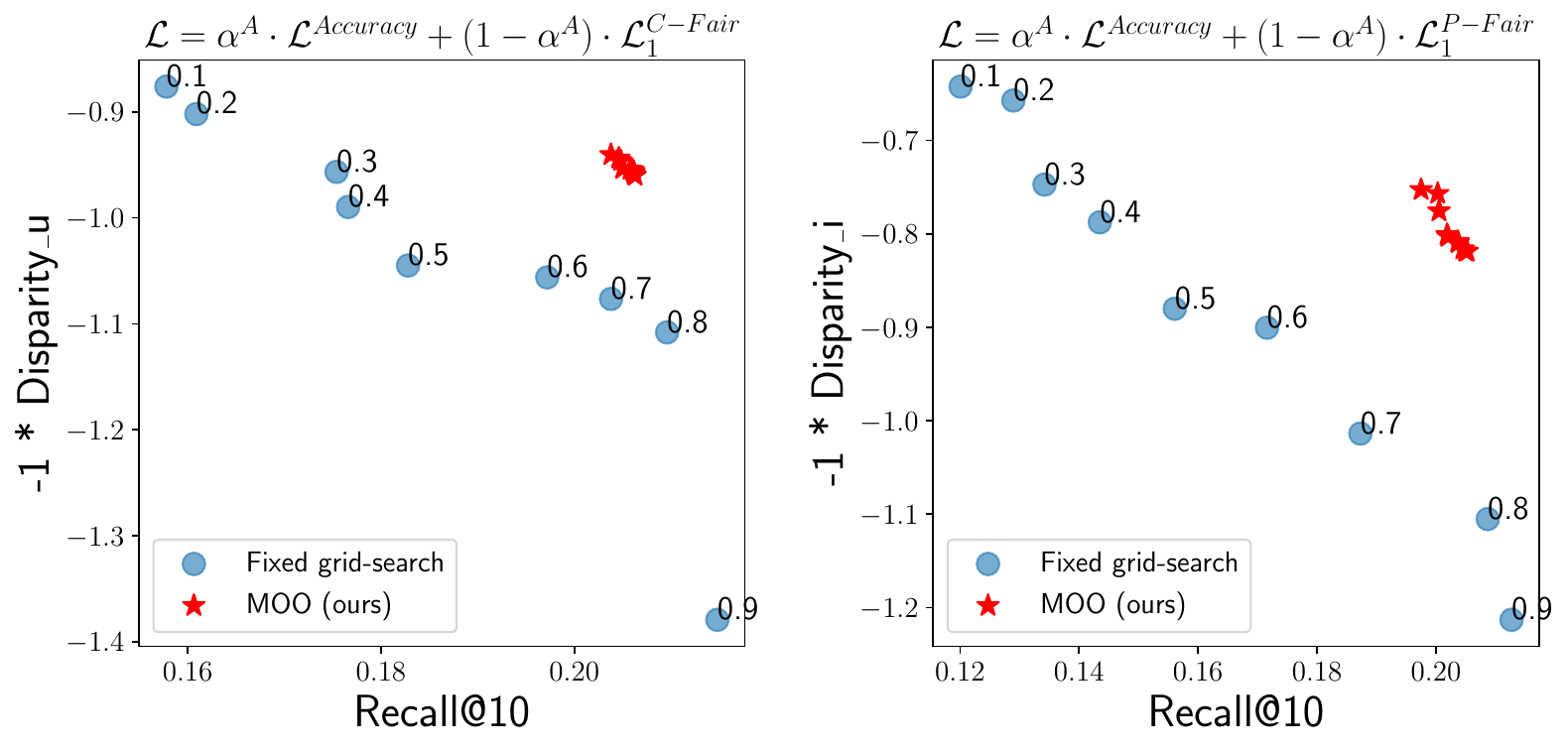}
\end{minipage}
}
\subfigure[\textbf{MovieLens1M}]{
\begin{minipage}[t]{1.0\linewidth}
\centering
\includegraphics[width=4.0in]{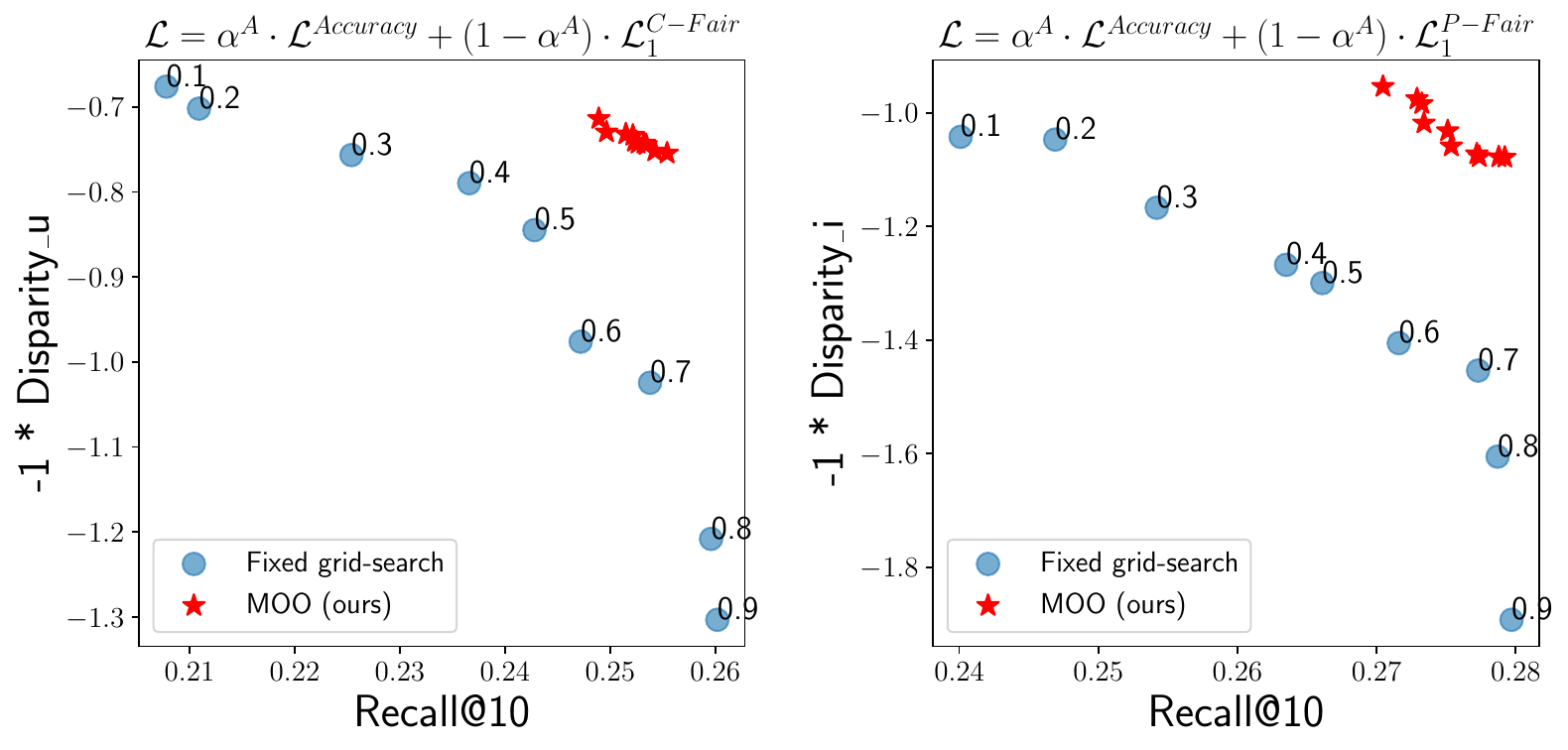}
\end{minipage}
}
\subfigure[\textbf{FM-reduced}]{
\begin{minipage}[t]{1.0\linewidth}
\centering
\includegraphics[width=4.0in]{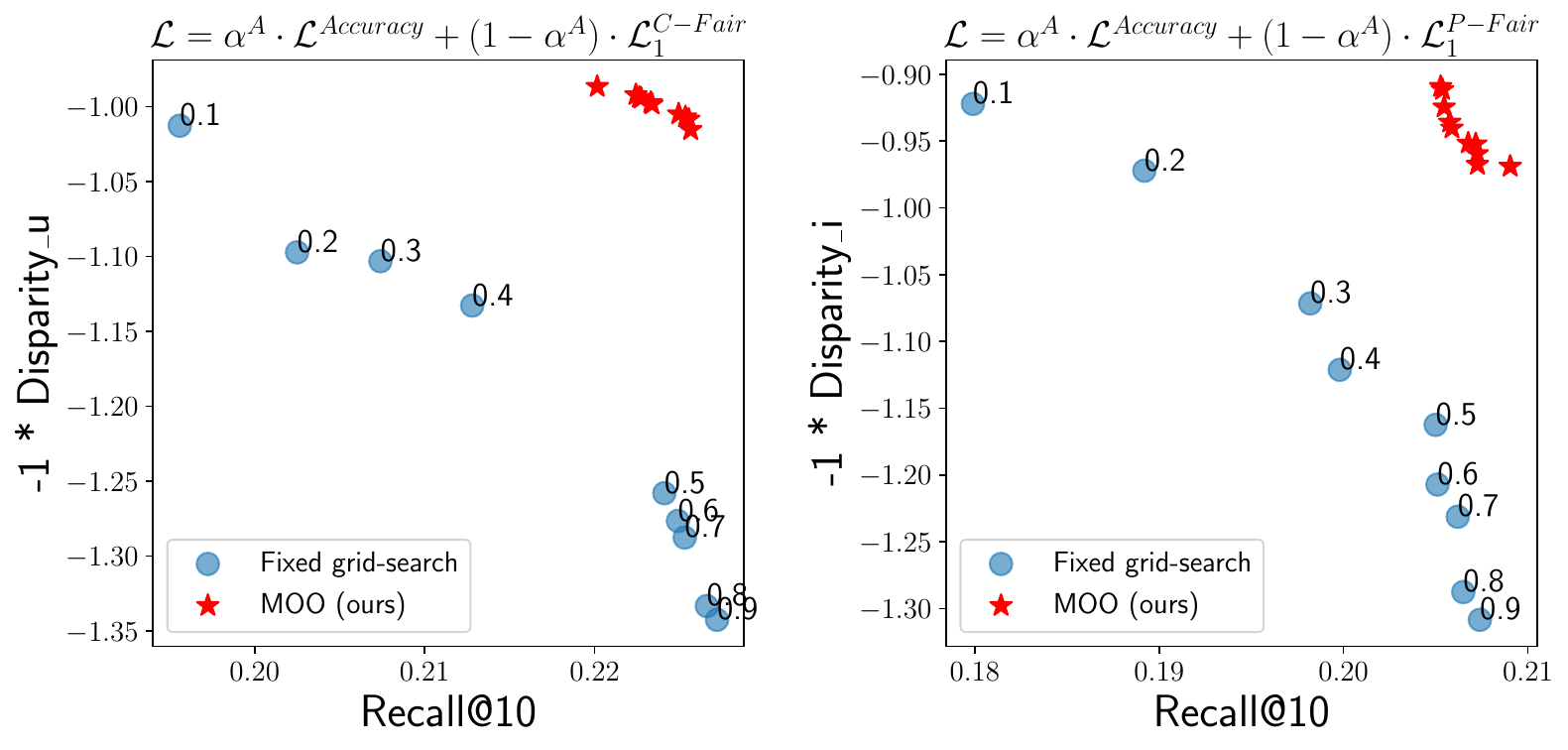}
\end{minipage}
}
\centering
\caption{The comparison between the MOO mechanism in our strategy versus the grid-search strategy on all datasets with the BPRMF backbone.}
\label{pic:moo_curve}
\end{figure}

\subsubsection{Comparison with Grid-search Strategy}
In order to demonstrate the effectiveness of the MOO mechanism in the Multi-FR, we conduct experiments to compare our model with the grid-search strategy, where scaling factors on the BPRMF objective and the fairness objective are manually set (the summation is 1). We only consider two-loss objectives for a convenient grid-search, which means we only add the fairness constraint on one side each time training with the BPR ranking loss (i.e., $\mathcal{L}=\alpha^A\cdot\mathcal{L}^{Accuracy}+(1-\alpha^A)\cdot \mathcal{L}_1^{C-Fair}$ or $\mathcal{L}=\alpha^A\cdot\mathcal{L}^{Accuracy}+(1-\alpha^A)\cdot \mathcal{L}_1^{P-Fair}$). The scatter plots are shown in Fig.~\ref{pic:moo_curve}. Each blue point indicates a grid-search solution averaged by 5 rounds where the value on the point is the weight $\alpha^A$ on the BPR loss. Each red point refers to one final Multi-FR solution selected through the strategy described in Section~\ref{sub:solution_selection} after running the model for 5 rounds. It is worthy to notice that all red points are Pareto optimal, which is theoretically guaranteed as demonstrated in Section~\ref{sec:multifr}. Thus, any red point cannot dominate other red points with respect to both the recommendation accuracy and the fairness. Here, the curve of the red points can be regarded as the Pareto frontier.

From the illustration in Fig.~\ref{pic:moo_curve}, we can observe that the MOO successfully balances the trade-off between the fairness and recommendation accuracy. The clear margin distance between the curve formed by the red points (Pareto frontier) and the curve formed by the blue points show the effectiveness of the MOO mechanism in our proposed Multi-FR.

\subsubsection{Training with Different Number of Constraints}
We investigate the empirical training efficiency by using a different number of fairness constraints in our model. We choose BPRMF as our base model to report the training efficiency. Each row in Table~\ref{tab:train_efficiency} indicates training with a different number of disparity objectives on the consumer side and the producer side. 
\new{Here the first and second fairness constraints on the consumer side represent the \textit{gender-based fairness} and \textit{age-based fairness}, respectively. The first and second fairness constraints on the producer side represent the \textit{popularity-based fairness} and \textit{genre-based fairness}, respectively.}
We observe that our proposed approach has reasonable training time, especially when the number of fairness constraints increases: the more number of constraints added, the less extra time the model needs. This shows the ability of our model to train multiple objectives simultaneously for multiple stakeholders in the real-world application.

\begin{table}[t]
\caption{The training efficiency comparison of different number of fairness constraints by using our model, \textit{Multi-FR}. The training time is reported in seconds. The backbone is chosen as the BPRMF.}
\renewcommand{\arraystretch}{1.6}
\vspace{1mm}
\label{tab:train_efficiency}
\centering
\begin{tabular}{lccc}
\toprule
Objective & \textit{ML100K} & \textit{ML1M} & \textit{FM-reduced}\\
\midrule
(1) $\alpha^A\cdot\mathcal{L}^{Accuracy}+\alpha_1^C\cdot \mathcal{L}_1^{C-Fair}$ & 676.4 & 11,059.2 & 69,438.3 \\
(2) $\alpha^A\cdot\mathcal{L}^{Accuracy}+\alpha_1^P\cdot \mathcal{L}_1^{P-Fair}$ & 363.8 & 8,945.1  & 54,281.8\\
(3) $\alpha^A\cdot\mathcal{L}^{Accuracy}+\sum\limits_{i=1}^2\alpha_i^C\cdot \mathcal{L}_i^{C-Fair}$ & 749.3 & 15,044.0  & 98,177.2\\
(4) $\alpha^A\cdot\mathcal{L}^{Accuracy}+\sum\limits_{i=1}^2\alpha_i^P\cdot \mathcal{L}_i^{P-Fair}$ & 512.5 & 12,684.2 & 90,864.5\\
(5) $\alpha^A\cdot\mathcal{L}^{Accuracy}+\alpha_1^C\cdot \mathcal{L}_1^{C-Fair}+\alpha_1^P\cdot \mathcal{L}_1^{P-Fair}$ & 912.7 & 19,560.7 & 102,232.1 \\
(6) $\alpha^A\cdot\mathcal{L}^{Accuracy}+\sum\limits_{i=1}^2\alpha_i^C\cdot \mathcal{L}_i^{C-Fair}+\alpha_1^P\cdot \mathcal{L}_1^{P-Fair}$ & 1,119.2 & 23,653.5 & 123,171.7\\
(7) $\alpha^A\cdot\mathcal{L}^{Accuracy}+\alpha_1^C\cdot \mathcal{L}_1^{C-Fair}+\sum\limits_{i=1}^2\alpha_i^P\cdot \mathcal{L}_i^{P-Fair}$ & 1,013.8 & 23,189.3 & 120,816.9\\
(8) $\alpha^A\cdot\mathcal{L}^{Accuracy}+\sum\limits_{i=1}^2\alpha_i^C\cdot \mathcal{L}_i^{C-Fair}+\sum\limits_{i=1}^2\alpha_i^P\cdot \mathcal{L}_i^{P-Fair}$ & 1,213.5 & 25,793.9 & 165,287.6\\
\bottomrule
\end{tabular}

\end{table}


\section{Conclusion and Discussion}
In this paper, we propose a multi-objective optimization framework, \textit{Multi-FR}, for the fairness-aware recommendation in multi-sided marketplaces, \new{where the final solution is guaranteed to be Pareto optimal.}
\new{To achieve fairness-aware recommendation, four fairness constraints are proposed within the multi-objective optimization framework.
We employ the smooth rank and stochastic ranking policy to make our fairness metrics differentiable, thus the fairness criteria can be optimized directly in an end-to-end way.}
Then, \textit{Multi-FR} applies the multi-gradient descent algorithm to generate a Pareto set, where the scaling factors on each objectives are adaptively learned through the Frank-Wolfe Solver without handcraft tuning. 
Finally, the \textit{Least Misery Strategy} is adopted to select the most proper solution from the generated Pareto set. 
Experimental results on three real-world datasets show that our method can constantly outperform various corresponding base architectures and state-of-the-art fairness recommendation/ranking methods. 
Extensive experiments on multiple evaluation metrics clearly validate that \textit{Multi-FR} can largely improve the recommendation fairness with only little drop in terms of the recommendation quality. 
Further analysis demonstrates the effectiveness of the MOO mechanism and the capability of \textit{Multi-FR} optimizing any number of fairness criteria for multiple stakeholders concurrently.

\new{There exist several extensions we intend to investigate as future work.
Firstly, we intend to collect the producer information for the datasets, such as film producers of movies, so that we can directly define and optimize producer-sided fairness, rather than making items act as a proxy of producers.
Secondly, our current definition of fairness only considers one demographic attribute at a time on the consumer side or the producer side. We intend to investigate how to ensure fairness for those people belonging to multiple demographic groups (e.g., “Black Women”).
Therefore, we need to consider how to model the fairness for multiple attributes (“color” and “gender” in this example) concurrently on one side. 
Thirdly, the Gumbel noise used to solve the non-differential problem when modeling the exposure fairness for producers is often independently and identically sampled. 
This is a common practice in community. However, we would like to study whether the model uncertainty can be integrated into this sampling procedure so that the final ranking positions based on the stochastic ranking policy can be estimated with a higher quality.}

\begin{acks}
We thank the constructive suggestions from anonymous reviewers. 
This work is supported by the Microsoft Research \& MILA (Quebec AI Institute) Collaboration Grant and the Start-up Grant (Grant \#: 9610564) of City University of Hong Kong. 
\end{acks}

\newpage
\bibliographystyle{ACM-Reference-Format}
\bibliography{reference.bib}

\end{document}